\documentclass[letterpaper, 11pt, leqno]{article}
\usepackage{paper,math}
\hypersetup{pdftitle={Modeling Migration-Induced Unemployment}}
\newcommand{\bib}{paper.bib}
\newcommand{\pdf}{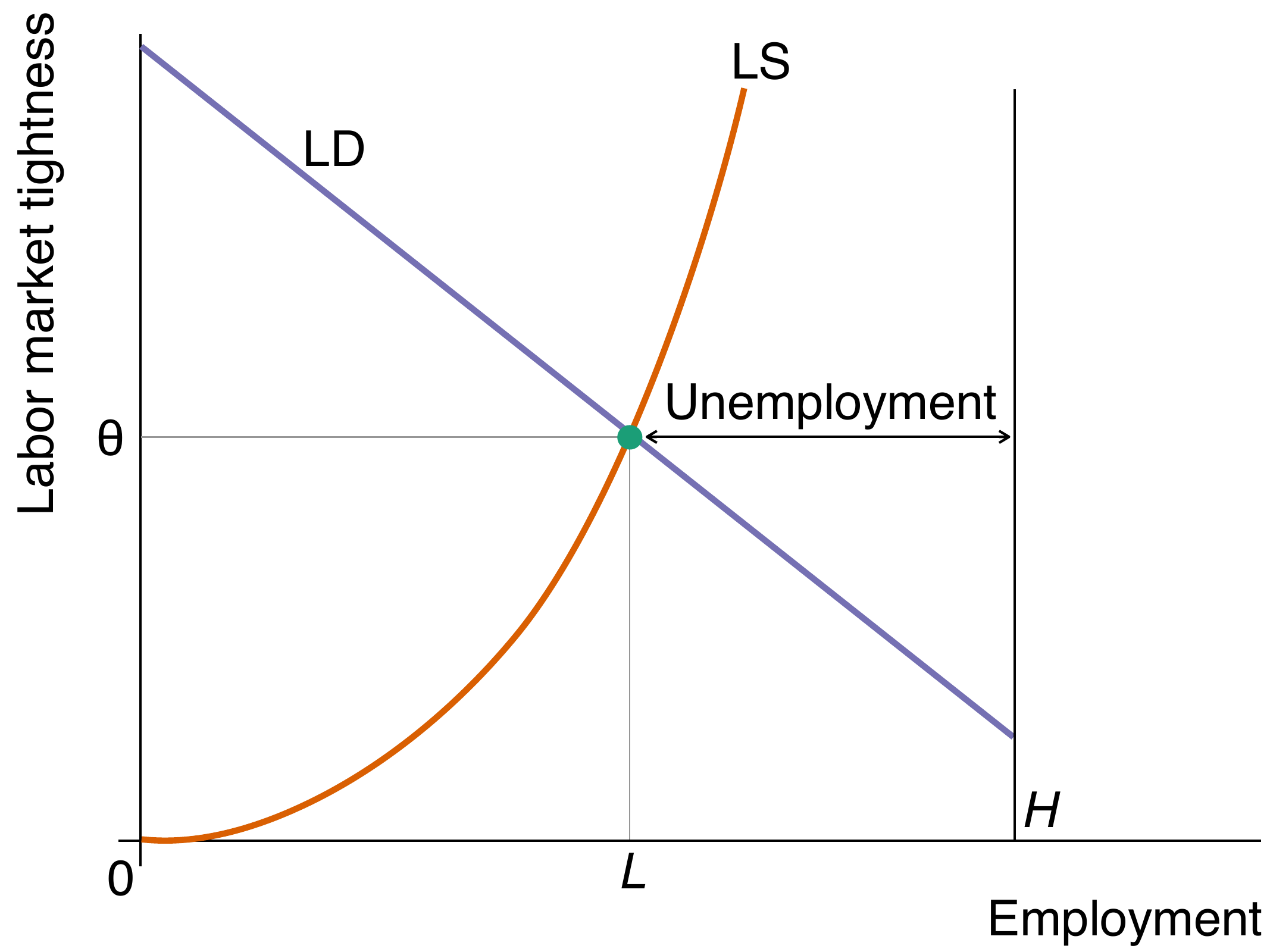}
\begin{document}

\begin{titlepage}
\title{Modeling Migration-Induced Unemployment}
\author{Pascal Michaillat
\thanks{University of California--Santa Cruz. I thank Avidit Acharya, Christoph Albert, Leah Boustan, Varanya Chaubey, Evgeniya Duzhak, Anthony Edo, Ahmet Gulek, Yagan Hazard, Akira Ishide, Gaurav Khanna, Claudio Labanca, Michael Lachanski, Ken Lee, Carlo Medici, Joan Monras, Hillel Rapoport, Gary Richardson, Emmanuel Saez, Paola Sapienza, Stefanie Stantcheva, Marco Tabellini, Jonathan Vogel, Romain Wacziarg, and Josef Zweimueller for helpful comments and discussions.}}
\date{September 2026}
\paperurl{https://pascalmichaillat.org/14/}
\maketitle

Immigration is often blamed for increasing unemployment among local workers. This sentiment is reflected in the rise of anti-immigration parties and policies in Western democracies. And in fact, numerous studies estimate that in the short run, the arrival of new workers in a labor market raises the unemployment rate of local workers. Yet, standard migration models inherently assume that migrants are absorbed into the labor market without affecting local unemployment. This paper presents a more general model of migration that allows for the possibility that not only the wages but also the unemployment rate of local workers may be affected by migration. The model blends a matching framework with job rationing. In it, the arrival of new workers raises the unemployment rate among local workers, particularly in a depressed labor market where job opportunities are limited. On the positive side, in-migration helps firms fill vacancies more easily, boosting their profits. The overall impact of in-migration on local welfare varies with labor market conditions: in-migration reduces welfare when the labor market is inefficiently slack but enhances it when the labor market is inefficiently tight. Finally, with high-skill and low-skill labor, in-migration always raises unemployment in the skill group that receives the migrants, while unemployment in the other group rises when the two skills are net substitutes in production and falls when they are net complements.

\end{titlepage}
\section{Introduction}
 
People are often worried about immigrants taking their jobs \citep{J01,S10}. Not surprisingly, then, immigration is one of people's chief policy concerns \citep{CDP12,LJ23}. In fact, since the 1970s, specifically anti-immigration political parties have appeared and started to grow in Western countries, now attracting 11\% of votes on average across these countries \citep[p.~22]{GMP22}. And far-right parties with anti-immigration platforms have become especially popular in countries where many immigrants live \citep[figure~1]{HWZ17}. 

People's concerns are consistent with evidence from several episodes during which the unemployment rate of local workers rose after the arrival of new workers in their labor market. On average across 11 migration experiments, the arrival of 100 new workers in a labor market pushed 37 local workers into unemployment (table~\ref{t:evidence}).

Yet, standard models of migration rule out the possibility that the unemployment rate of local workers may be affected by newcomers. When migration takes place in a Walrasian labor market, there is no unemployment at all: anybody who wants to work can work. In a Diamond-Mortensen-Pissarides (DMP) labor market, there is some unemployment, but labor demand is perfectly elastic. Thus, newcomers are absorbed into the labor market without affecting other workers (appendix~\ref{a:dmp}). Hence, the standard models do not capture the phenomenon that newly arrived workers compete for jobs with local workers and raise their unemployment rate. 

Because they rule out the possibility that the local unemployment rate may be affected by migration, standard models cannot describe and explain the short-run response of unemployment to migration. They cannot explain through which channels and under which conditions migration may induce unemployment. They cannot predict when such a response is likely to be severe and when it is likely to be mild. And they cannot be used to design compelling migration policies because they rule out what is in the public's mind one of the most deleterious effects of migration. In particular, they cannot say whether migration policies should respond to the business cycle or the state of the local economy because they assume that jobs are always plentiful.

This paper therefore presents a model that allows for migration-induced unemployment. In the labor market model from \citet{M12}, which adds job rationing to the DMP model, the entry of migrants reduces labor market tightness and thus the job-finding rate of local workers. As a result, the arrival of migrants increases the unemployment rate of local workers---who might naturally feel that migrants take their jobs away. 

The mechanism is as follows. In normal times, the number of jobs available is somewhat limited. When new workers enter the labor force, the number of jobseekers increases, so labor market tightness drops. This makes hiring more attractive to firms, so they hire some more workers---some local and some newly arrived. While this additional hiring boosts labor market tightness, tightness remains below its initial level because the number of jobs available is not sufficient to absorb all newcomers. Since labor market tightness is lower, the prevailing job-finding rate is lower, so the unemployment rate goes up. That is, fewer local workers hold jobs.

In bad times, the lack of jobs is more stringent, so firms absorb fewer newcomers. More newcomers remain jobless, which increases the competition for jobs. As a result, local workers are more negatively affected by in-migration in bad times. Formally, the elasticity of local employment with respect to in-migration is more negative in bad times. This explains why the backlash against immigration is stronger when the unemployment rate is elevated \citep{HWZ17}. 

While in-migration hurts workers, it helps firms. This is because in-migration reduces labor market tightness, so it makes it easier for firms to recruit workers. Easier recruiting in turn improves firm profits.

The overall effect of in-migration on local welfare---the sum of local labor income and local firm profits---depends on the health of the labor market. When the labor market is inefficiently slack, allowing in-migration reduces local welfare. But in-migration improves local welfare when the labor market is inefficiently tight. In that case, the local labor income lost from in-migration is more than offset by the increase in firm profits. 

In the model calibrated to the US labor market, the amount of migration-induced unemployment is substantial and commensurate to the amount observed in migration experiments. For instance, 33 local workers become unemployed when 100 migrants arrive in their labor market and labor market tightness is at its efficient level of 1. When labor market tightness drops to 0.5 the number of local workers who become unemployed rises to 49; when labor market tightness climbs to 2, the number of local workers who become unemployed falls to 20.

An extension of the model with low-skill and high-skill workers who contribute separately to production shows how migration into one skill group propagates into the other skill group through firms' substitution between skills. The impact of migration with heterogeneous labor is determined by the elasticity of substitution between skills relative to the inverse of the diminishing-returns parameter. If skills are substitutable enough, workers of both skills face a higher unemployment after in-migration. If skills are complementary enough, only the directly affected skill faces more unemployment; the other one enjoys less unemployment.

In this paper the focus is on the short-term effects of in-migration. In the long run, immigration may have positive effects for the aggregate economy, in particular through increased innovation.\footnote{See for instance \citet{HG10}, \citet{MVW14}, \citet{SNQ20}, \citet{BDJ22}, and \citet{TCB26}.} But political backlash is typically generated by short-run and local effects. The paper explains how immigration can generate unemployment in the short run in local labor markets. It therefore helps to understand public anxieties about immigration, which often center on the idea that immigrants take jobs away from local workers. In that, it adds to existing explanations for anti-immigrant sentiments, such as racism and xenophobia \citep{J21}.

This paper's mechanism for inducing local unemployment differs from, but complements, the mechanism in \citet{A21}. Both papers use a matching model, but they isolate different channels through which immigration can raise local unemployment. In \citet{A21}, migrants and locals are perfect substitutes in production but have different reservation wages. Furthermore, hiring is nonrandom: firms collect multiple applications and select the worker willing to accept the lowest wage. Immigration then has two opposing effects on local employment: a job-creation effect, because lower average wage costs induce firms to post more vacancies, and a job-competition effect, because firms prefer to hire cheaper migrant applicants over otherwise identical local applicants. Local unemployment rises when the job-competition effect dominates the job-creation effect, and it declines otherwise. In simulations of the US labor market, \citeauthor{A21} finds that undocumented immigrants (who accept much lower wages) reduce local unemployment via job creation while documented immigrants (who only accept slightly lower wages) increase local unemployment via job competition. By contrast, the mechanism in this paper does not require any heterogeneity between migrants and locals: even when both groups are paid the same, immigration lowers labor market tightness because labor demand does not fully scale up in the short run, which reduces the job-finding rate of local jobseekers and raises local unemployment. The two mechanisms are therefore complementary: the effect of job rationing could be reinforced by selective hiring when migrants accept lower wages than locals.

\section{Evidence of migration-induced unemployment}\label{s:evidence}

This section presents evidence of migration-induced unemployment from natural and quasi experiments.\footnote{Appendix~\ref{a:other} provides additional nonexperimental evidence that the arrival of immigrants on a labor market reduces the number of local workers who are employed.} The experiments correspond to events in which newcomers arrive in local labor markets for plausibly exogenous reasons, often because they were forced to migrate by external events such as wars or political crises or economic upheavals.\footnote{As \citet[p.~25]{P16} notes, these experiments naturally have some limitations. For instance, the people who migrate during these events may be different from typical migrants. Moreover, even if immigrants arrive in a country for non-economic reasons, where they settle might be influenced by local economic conditions, possibly threatening the exogeneity of their arrival. As long as immigrants target places with low unemployment rates or decreasing unemployment rates, however, the endogeneity bias only dampens the reported results, so the true increase in local unemployment caused by in-migration would be larger.} The experiments occurred across different countries and historical periods, and they involved both international and domestic migrations, but they paint a consistent pattern: an influx of new workers into a local labor market raises the unemployment rate among local workers. The evidence is summarized in table~\ref{t:evidence}.

\begin{sidewaystable}[p]
\caption{Migration-induced unemployment across countries and periods}
\begin{tabular*}{\textwidth}{@{\extracolsep{\fill}}p{2.9cm}cccc}\toprule
Unemployed locals per 100 arrivals & Country & Period & Event & Reference\\
\midrule
\multicolumn{5}{c}{A. Domestic migration}\\
21 & United States & 1935--1940 & Great Depression & \citet{BFK10} \\
\midrule
\multicolumn{5}{c}{B. Return migration}\\
20 & France & 1962 & Algerian War of Independence & \citet{H92} \\
22 & Portugal & 1974--1977 & Independence of Mozambique \& Angola & \citet{M17a} \\
31 & Germany & 1996--2001 & Fall of Berlin Wall & \citet{G12} \\
\midrule
\multicolumn{5}{c}{C. International migration}\\
24--27 & Algeria $\to$ France & 1962 & Algerian War of Independence & \citet{BM17} \\
15 & Cuba $\to$ United States & 1980--1981 & Mariel Boatlift & \citet{C90} \\
66--77 & Czech Republic $\to$ Germany & 1991--1993 & Fall of Berlin Wall & \citet{DSS17} \\
30--40 & Eastern Europe $\to$ West Germany & 1987--2001 & Fall of Berlin Wall & \citet{DOP10} \\
35--83 & Yugoslavia $\to$ Europe & 1983--1999 & Yugoslav Wars & \citet{AK03} \\
21--47 & Yugoslavia $\to$ Europe & 1991--2001 & Yugoslav Wars & \citet{BM17} \\
63--80 & North Africa $\to$ Italy & 2011 & Arab Spring & \citet{L20} \\
\bottomrule
\multicolumn{5}{c}{Summary statistics for the number of unemployed locals per 100 arrivals}\\
 Mean: 37 & Median: 31 & Minimum: 15 & Maximum: 72 & Standard deviation: 20 \\
\bottomrule\end{tabular*}
\note{The evidence described in the table is obtained in Section~\ref{s:evidence}. For studies that provide a range of estimates, the summary statistics use the midpoint of the range.}
\label{t:evidence}\end{sidewaystable}

\subsection{US Great Depression}

\citet{BFK10} study the effect of internal migration in the United States during the Great Depression. Just like native workers protest immigration from abroad, locals in areas where the depression was less severe protested against the arrival of migrants from other less fortunate regions. Locals accused newcomers of taking jobs away from them. For example, Californians tried to scare possible migrants away from the state. A billboard in Oklahoma carried the message ``NO JOBS in California / If YOU are looking for work—KEEP OUT / 6 men for every job / No state relief available for non-residents'' \citep[p.~720]{BFK10}.

To assess the impact of internal migrants on existing residents, \citet{BFK10} use variation in the generosity of New Deal programs and extreme weather events to instrument for migrant flows to and from US cities. They find that in-migration had little effect on the hourly earnings of existing residents, but it prompted some residents to move away from their city and others to lose weeks of work or access to relief jobs. They estimate that for every 100 arrivals, 21 locals were displaced from relief jobs. An additional 19 local workers were forced to shift from full-time to part-time work. And a further 19 residents moved out to other cities. They also find that, just as in-migration diminished the work opportunities of local workers, out-migration improved local opportunities symmetrically.

\subsection{Algerian War of Independence}

\citet{H92} examines the repatriation to France of Algerians of European origin following the independence of Algeria in 1962. \citet[p.~566]{H92} finds that the arrival of 100 repatriates in the labor force pushed 20 natives into unemployment.

\citet{BM17} look at Algerian refugees who fled Algeria for France at the same time as European repatriates. They aim to estimate the impact of immigration in individual region-education cells of the French labor market. For these refugees, \citet[table~10]{BM17} find larger effects than for the repatriates: 24--27 natives were pushed into unemployment for every 100 refugees in any region-education cell.

\subsection{Independence of Mozambique and Angola}

\citet{CL96} and \citet{M17a} study the return of half a million workers from Mozambique and Angola to Portugal in 1974--1976, following the independence of Mozambique and Angola. This return led to a 15\% increase in the Portuguese civilian labor force. \citet[table~2]{CL96} find that in 1981, returnees faced an unemployment rate of 14\%, much higher than the unemployment rate of 6\% faced by locals. This elevated unemployment rate indicates that it was difficult for returnees to get absorbed by the Portuguese labor market. Moreover, \citet[p.~248]{M17a} finds that in 1977, after the wave of migration subsided, the Portuguese unemployment rate was 3.3pp higher than in a counterfactual outcome. This means that for each 100 returnees into the labor force, $3.3/15 \times 100 = 22$ additional local workers were facing unemployment in 1977.

\citet[p.~242]{M17a} also reports that Portuguese workers were concerned about the return of workers from the African colonies, especially because the Portuguese economy was slumping at that time. The returnees were accused of ``stealing housing and jobs'' from Portuguese residents.

\subsection{Mariel Boatlift}

Interestingly, the results from \citet{C90}'s famous study are not inconsistent with migration-induced unemployment. He studies the impact of the Cuban immigrants from the Mariel Boatlift on the Miami labor market in the 1980s. A first, well-known finding from the study is that ``the Mariel immigration had essentially no effect on the wages or employment outcomes of non-Cuban workers in the Miami labor market'' \citep[p.~255]{C90}. A second finding is that ``perhaps even more surprising, the Mariel immigration had no strong effect on the wages of other Cubans'' \citep[p.~255]{C90}.

A third, less-known finding is that the unemployment rate for Cuban workers increased drastically: ``Unlike the situation for whites and blacks, there was a sizable increase in Cuban unemployment rates in Miami following the Mariel immigration. Cuban unemployment rates were roughly 3 percentage points higher during 1980--81 than would have been expected on the basis of earlier (and later) patterns'' \citep[p.~251]{C90}. That the Cuban workers faced a higher unemployment rate indicates that the labor market could not absorb all new arrivals, and that jobs were somewhat rationed. The boatlift raised the Cuban labor force in Miami by 20\% \citep[p.~246]{C90}. So for each 100 new arrivals into Miami's Cuban labor force, $3/20 \times 100= 15$ local Cubans were pushed into unemployment.

Relatedly, \citet[figure~5]{ABC21} compute the response of labor market tightness in Miami upon the arrival of Cuban immigrants from the Mariel Boatlift in the 1980s. Relative to the tightness in comparable cities, they find that Miami tightness fell by 40\% after the Mariel Boatlift. This implies that it became much harder for local workers to find jobs after the boatlift.

\subsection{Fall of the Berlin Wall}

\citet{G12} studies the return of 2.8 million ethnic Germans to Germany following the fall of the Berlin Wall. He estimates that for every 100 immigrants that found a job, 31 locals became unemployed, with no effect on wages. \citet{CV08} finds a similar number in the same empirical context.

\citet[pp. 462, 475, 476]{DSS17} estimate the impact of Czech commuters who were allowed to work in German border towns in 1991--1993, just after the fall of the Iron Curtain. They find evidence of migration-induced unemployment again. For each 100 commuters that found a job, 66--77 German workers were relegated to unemployment. Some further 16--27 German workers relocated to towns further away from the border, where Czech commuters did not compete for jobs.

Similarly, \citet[p.~559]{DOP10} estimate how West German workers were impacted by the arrival of ethnic Germans, East Germans, and foreigners into Germany in 1987--2001. They find that the arrival of new immigrants did not affect natives much, but it negatively affected previous immigrants, who were already working in West Germany. For 100 new immigrants who found a job, 30--40 previous immigrants were pushed into unemployment.

\subsection{Yugoslav Wars}

\citet[pp.~F318, F322]{AK03} look at migration from the former Yugoslavia into other European countries in the 1990s, because of the Yugoslav Wars. With an OLS specification, they find that the entry of 100 refugees in the labor force pushed 35 native workers into unemployment. With an IV specification, they find that the entry of 100 refugees in the labor force pushed 83 native workers into unemployment.

\citet[table~13]{BM17} revisit the evidence using a region-education-cell strategy. They observe again migration-induced unemployment, although their results are weaker. With an OLS specification, they find that 21 natives were pushed into unemployment when 100 refugees entered a region-education cell. With an IV specification, the number of natives pushed into unemployment rises to 47.

\subsection{Arab Spring}

Finally, \citet{L20} examines the arrival of Arab Spring refugees into the Italian labor market in 2011. He finds that for every 100 refugees employed, 63--80 Italian workers became unemployed.

\subsection{Discussion}\label{s:discussion}

This section shows that in-migration raises local unemployment in many settings. On average across the 11 migration experiments, 100 new arrivals in a labor market push 37 local workers into unemployment (table~\ref{t:evidence}). The amount of migration-induced unemployment varies substantially across experiments. The number of local workers pushed into unemployment ranges from 15 to 72 per 100 arrivals, with a standard deviation of 20.

Yet, academic economists tend to reject the notion that immigration negatively affects local workers. This has created a puzzle, as many people believe that immigration hurts them. In Europe, \citet[p.~78]{CDP12} observe that:
\begin{quote}
There is strong public opposition to increased immigration throughout Europe. Given the modest economic impacts of immigration estimated in most studies, the depth of anti-immigrant sentiment is puzzling. 
\end{quote}
In the United States, \citet[p.~302]{FHK06} add that:
\begin{quote}
One of the central questions in the debate over immigration policy is whether immigrants adversely affect labor market outcomes for natives. Some Americans believe they do, worrying that immigrants take jobs away from native workers. Most of the empirical evidence produced by economists, however, does not support these concerns.
\end{quote}

The general perception by academic economists that immigration only has ``modest'' effects on local workers might be an indirect consequence of not having models to think about the effect of migration on unemployment. The adverse impact of immigration on unemployment is not emphasized in economic research---but it is often there, offering a resolution to the puzzle. If such a phenomenon was present in standard models, researchers might have paid more attention to the empirical findings in existing studies. Indeed, one of the main functions of models is to guide and stimulate empirical work about the phenomena that they describe \citep{K57}. 

The absence of migration-induced unemployment in existing migration models limits the scope of empirical inquiry about migration. A striking example of such limitation appears in a seminal paper by \citet{SS01}. The paper documents people's perceptions about immigration. \citet[footnote~8]{SS01} candidly admit that they do not examine people's perception of job stealing because such perceptions would be inconsistent with the effects of immigration in standard models:
\begin{quote}
The 1992 National Election Studies survey asked other questions about immigration that we do not analyze. For example, respondents were asked whether they think Asians or Hispanics ``take jobs away from people already here.'' We do not focus on this question because its responses cannot clearly distinguish among our three competing economic models. All our models assume full employment, so no natives could have jobs ``taken away'' by immigrants.
\end{quote}

The response of respondents in the 1992 National Election Studies is presented in table~\ref{t:stealing}. More than 80\% of respondents are indeed worried that Hispanic and Asian immigrants take jobs away from them. It is quite possible that the existence of migration-induced unemployment has not received the attention it deserves just because standard models do not feature it.

\begin{table}[t]
\caption{Migration-induced unemployment in popular perceptions}
\begin{tabular*}{\textwidth}{@{\extracolsep{\fill}}p{6cm}*{4}{c}}
\toprule
 & \multicolumn{4}{c}{How likely is it?}\\
\cmidrule{2-5}
The growing number of these immigrants takes jobs away from people already here. & Extremely & Very & Somewhat & Not at all \\
\midrule
Hispanics 	& 20\% & 29\% & 38\% & 13\% \\
Asians	 	& 19\% & 30\% & 37\% & 13\% \\
\bottomrule\end{tabular*}
\note[Source]{1992 American National Election Studies survey. The responses to the question about Hispanic workers (question Q4c) can be accessed at \url{https://electionstudies.org/data-tools/anes-variable/variable.html?year=1992&variable=V926238}. The responses to the question about Asian workers (question Q5c) can be accessed at \url{https://electionstudies.org/data-tools/anes-variable/variable.html?year=1992&variable=V926241}.}
\label{t:stealing}\end{table}

\section{Model of the labor market}\label{s:model}

To formulate a theory of migration that makes sense of the evidence presented in section~\ref{s:evidence}, I use a model in which not only wages but also labor market tightness might respond to migration. The model, developed by \citet{M12}, inserts job rationing into the DMP matching model. This is done by assuming that the production function is concave instead of linear in labor, and that wages are determined by a rigid wage norm instead of Nash bargaining. Introducing job rationing is required to generate migration-induced unemployment.

\subsection{Assumptions}

The model is dynamic and set in continuous time. The labor market is composed of a labor force of size $H>0$ and a representative firm. Unemployed workers and vacant jobs match at a rate given by a Cobb-Douglas matching function:
\begin{equation}
m(U,V) = \m U^{\h} V^{1-\h},
\label{e:matching}\end{equation} 
where $U$ is the number of job seekers, $V$ is the number of job vacancies, $\m>0$ is the matching efficacy, and $\h\in (0,1)$ is the matching elasticity.

The state of the labor market is captured by labor market tightness:
\begin{equation*}
\t = \frac{V}{U}.
\end{equation*}
Labor market tightness plays such an important role because it determines the rate at which unemployed workers match with vacant jobs,
\begin{equation}
f(\t) = \frac{m(U,V)}{U} = \m \t^{1-\h},
\label{e:f}\end{equation}
and the rate at which vacant jobs are filled with unemployed workers,
\begin{equation}
q(\t)= \frac{m(U,V)}{V} = \m \t^{-\h}.
\label{e:q}\end{equation}
The job-finding rate $f(\t)$ is increasing in tightness $\t$, and the job-filling rate $q(\t)$ is decreasing in tightness $\t$. Hence, when tightness is higher, it is easier for workers to find jobs but harder for the firm to find workers. The representative firm takes labor market tightness as given.

The firm has a concave production function: 
\begin{equation}
y(P) = a P^{1-\a},
\label{e:production}\end{equation} 
where $a>0$ governs labor productivity, $P$ denotes the number of producers in the firm, and $\a \in (0,1)$ governs diminishing returns to labor. 

Why might the production function exhibit diminishing instead of constant returns to labor? A basic reason is that in the short run, capital and other factors of production are fixed. For instance, if the long-run production function is given by a Cobb-Douglas function of labor and capital, $y(K,P) = b K^{\a} P^{1-\a}$, but capital $K$ is fixed in the short run, then in the short run the production function takes the form \eqref{e:production}.\footnote{In their immigration models, \citet[equation~(1)]{OP12} assume such a Cobb-Douglas production function, and \citet[section~2.1]{B99} assumes that the capital stock is fixed. In the longer run, the impact of migration on the labor market might diminish as firms adjust their capital stock to the labor stock \citep{C12a}.} 

The firm also incurs a recruiting cost of $\k > 0$ recruiters per vacancy and faces a job-separation rate $\l > 0$. The total number of recruiters in the firm is $R = \k V$ and the total number of workers is $L = R + P$.

Finally, as in any matching model, wages are determined in a situation of bilateral monopoly, so it is a wage norm and not an auctioneer that determines wages \citep{HM87,H05}. For now, for simplicity, I assume that real wages are fixed at $w>0$. In particular, wages do not respond to migration. The aim of the assumption is to concentrate on the response of tightness to migration---which is the key innovation of the paper. 

Once the response of tightness is well understood, however, I will generalize the wage norm and allow it to respond to migration (section~\ref{s:wage}). The qualitative results will be the same whether wages are completely fixed or partially rigid, as long as wages are not completely flexible.

\subsection{Labor supply}

In the model, the employment level is given by the following law of motion:
\begin{equation}
\dot{L}(t) = f(\t) U(t) - \l L(t).
\label{e:ldot}\end{equation} 
That is, the employment level increases over time ($\dot{L}>0$) if more jobseekers find jobs than employed workers lose their jobs ($f(\t) U > \l L$). Conversely, employment decreases over time if more employed workers lose their jobs than jobseekers find jobs.

Since $U(t) = H - L(t)$, the law of motion \eqref{e:ldot} can be rewritten as the following differential equation:
\begin{equation*}
\dot{L}(t) = f(\t) H - \bs{\l+f(\t)} L(t).
\end{equation*} 
The critical point of this differential equation is
\begin{equation}
L = \frac{f(\t)}{\l+f(\t)} \cdot H.
\label{e:l}\end{equation}
This positive relationship between employment and tightness is the locus of employment and tightness such that the number of new employment relationships created at any point in time equals the number of relationships dissolved at any point in time. It is the locus of points such that the employment level remains constant over time and labor market flows are balanced.

In the US labor market, the employment levels given by equations~\eqref{e:ldot} and~\eqref{e:l} are indistinguishable.\footnote{See for instance \citet[figure~1]{H05a} and \citet[figure~A3]{MS21b}.} I therefore assume that at the time scale of the model, labor market flows are always balanced, and the employment level is given by equation~\eqref{e:l} at all times. The employment level consistent with balanced flows is the labor supply:
\begin{equation}
L^s(\t) = \frac{f(\t)}{\l+f(\t)} \cdot H.
\label{e:ls}\end{equation}
The labor supply links employment to tightness at any point in time. 

The labor supply also relates the unemployment rate to tightness. The unemployment rate is $u = 1 - L^{s}/H$ so
\begin{equation}
u(\t) = \frac{\l}{\l+f(\t)}.
\label{e:u}\end{equation}

\subsection{Recruiter-producer ratio}

Because it takes time to fill vacancies and each vacancy requires the attention of a recruiter, the firm allocates a share of its workforce to recruiting. And because the model is cast on a time scale where labor market flows are balanced, the firm posts vacancies to maintain its size at a desirable level. That is, it posts $V$ vacancies so the $q(\t)V$ new hires just replace the $\l L$ workers who left. 

Multiplying $\l L = q(\t)V$ by the recruiting cost $\k$, and using $L = R+P$ and $R = \k V$, yields $\l \k (P+R) = q(\t) R$. Dividing both sides by $R$ then gives $\l \k (1+\tau(\t)^{-1}) = q(\t)$, where $\tau(\t) = R/P$ is the recruiter-producer ratio. The recruiter-producer ratio is therefore given by
\begin{equation}
\tau(\t) = \frac{\l \k}{q(\t)-\l \k}.
\label{e:tau}\end{equation}
This means that
\begin{equation}
1+\tau(\t) = \frac{q(\t)}{q(\t)-\l \k}.
\label{e:1+tau}\end{equation}
The recruiter-producer ratio $\tau(\t)$ is positive and increasing on $(0,\bar{\t})$, where $\bar{\t}$ is defined by $q(\bar{\t}) = \l \k $; furthermore, $\lim_{\t\to 0} \tau(\t) = 0$ and $\lim_{\t\to \bar{\t}} \tau(\t) = + \infty$.

The wedge $1+\tau(\t)$ plays an important role in the analysis because it determines the gap between the numbers of employees and producers in the workforce: 
\begin{equation}
L = P + R = \bs{1+\tau(\t)} P.
\label{e:lnr}\end{equation}

\subsection{Labor demand}

The firm operates within the balanced-flow paradigm. By choosing how many vacancies to post, it determines its workforce, which it chooses to maximize real flow profits: 
\begin{equation*}
y(P) - wL = y(P) - \bs{1+\tau(\t)} w P. 
\end{equation*}
The first-order condition of the firm's maximization problem is $y'(P) - \bs{1+\tau(\t)} w = 0$, or
\begin{equation}
(1-\a) a P^{-\a} =\bs{1+\tau(\t)} w.
\label{e:foc}\end{equation}
This condition says that at the optimum, the marginal product of labor equals the marginal cost of labor.

With \eqref{e:lnr}, the first-order condition \eqref{e:foc} becomes
\begin{equation*}
(1-\a) a \bs{1+\tau(\t)}^{\a}{L}^{-\a} = \bs{1+\tau(\t)} w,
\end{equation*}
which then yields the firm's labor demand:
\begin{equation}
L^d(\t) = \bc{\frac{(1-\a) a}{\bs{1+\tau(\t)}^{1-\a} w}}^{\frac{1}{\a}}.
\label{e:ld}\end{equation}
The labor demand gives the firm's employment level for any tightness and productivity. There is job rationing in the model because the labor demand is decreasing with tightness. The labor demand fluctuates in response to productivity shocks over the business cycle.

\subsection{Solution of the model}

In the model, the firm maximizes profits, and employment is determined by the matching and separation processes. These requirements impose that labor market tightness $\t$ equalizes labor demand and supply:
\begin{equation}
L^d(\t) = L^s(\t).
\label{e:solution}\end{equation}

Given the properties of the labor demand and supply curves, it is clear that the model admits a unique solution. (The formal proof is simple and relegated to appendix~\ref{a:solution}). The solution of the model is illustrated in figure~\ref{f:solution}. The labor demand curve is downward sloping because a tighter labor market makes it harder for firms to recruit and reduces their demand for labor. The labor supply curve is upward sloping because in a tighter labor market, jobseekers find jobs faster so more people end up employed. The tightness and employment that solve the model are read at the intersection of the two curves. Unlike in a Walrasian model, there is always some unemployment here: the number of unemployed workers is given by the horizontal distance from the solution to the labor force.

\begin{figure}[p]
\subcaptionbox{Solution of the model \label{f:solution}}{\normalgraphic[1]{\pdf}}\hfill
\subcaptionbox{Effects of in-migration\label{f:inmigration}}{\normalgraphic[2]{\pdf}}\\
\subcaptionbox{In-migration when wages fall\label{f:wage}}{\normalgraphic[3]{\pdf}}\hfill
\subcaptionbox{Drop in labor productivity\label{f:shock}}{\normalgraphic[4]{\pdf}}\\
\subcaptionbox{In-migration in good times\label{f:goodtimes}}{\normalgraphic[5]{\pdf}}\hfill
\subcaptionbox{In-migration in bad times\label{f:badtimes}}{\normalgraphic[6]{\pdf}}
\caption{Description of the model and the effects of in-migration}
\note{The labor supply curve (LS) is given by \eqref{e:ls}. The labor demand curve (LD) is given by \eqref{e:ld}. In panels B, C, E, and F, the dashed lines represent the curves before migration, the solid lines represent the curves after migration, and the arrows show the decrease in local employment induced by in-migration. In panel D, the dashed line is the labor demand curve with high labor productivity and the solid line is the labor demand curve after the drop in productivity. Panel A illustrates the solution of the model. Panel B illustrates the results from proposition~\ref{p:migration}. Panel C illustrates the results from proposition~\ref{p:wage}. Panel D illustrates the results from proposition~\ref{p:businesscycle}. Panels E and F illustrate the results from proposition~\ref{p:stealing}.}
\label{f:model}\end{figure}

\section{Impact of in-migration on the labor market}\label{s:inmigration}

This section describes how a wave of in-migration affects the labor market. It shows that migration alters labor market tightness and thus the unemployment rate of local workers. 

\subsection{Modeling in-migration}

Migration leads to a change in the number of workers in the labor force. To model migration, I therefore set the labor force to $H \cdot I$, where $I>0$ is the migration factor. The case $I = 1$ represents no migration; $I>1$ represents in-migration; and $I<1$ represents out-migration. To simplify some of the calculations, I will sometimes use the migration rate $i$, defined by $i = I - 1$. In-migration corresponds to $i>0$ and out-migration to $i<0$. The migration rate measures the amount of migration as a share of the local labor force.

With migration, the labor supply \eqref{e:ls} becomes
\begin{equation}
L^s(\t,I) = \frac{f(\t)}{\l+f(\t)} \cdot H \cdot I.
\label{e:lsi}\end{equation}
and labor market tightness $\t$ is given by
\begin{equation}
L^d(\t) = L^s(\t,I).
\label{e:solutioni}\end{equation}
Now, the supply-equals-demand condition implicitly defines tightness as a function $\t(I)$ of the migration factor. The unemployment rate also becomes a function of the migration factor: $u(\t(I))$, where $u(\t)$ is given by \eqref{e:u}. The aggregate number of employed workers---both local and migrant---is read off the labor demand: $L(I) = L^d(\t(I))$. The number of local workers who are employed is 
\begin{equation}
N(I) = \bs{1 - u(\t(I))} \cdot H.
\label{e:ni}\end{equation}
Indeed, there are $H$ local workers in the labor force, and they face an unemployment rate $u(\t(I))$. The number of migrant workers who are employed is $\bs{1 - u(\t(I))} \cdot H \cdot i$.

By modeling in-migration as an increase in the labor force, I implicitly assume that local and migrant workers are the same: they are hired through the same channels, have the same productivity, and command the same wage.\footnote{\citet[section~2.1]{B99} similarly studies migration with homogeneous labor.} This assumption is sometimes unsatisfactory. For instance, in the case of international migration to the United States, immigrants and natives generally differ in terms of education and experience \citep{B03,OP12}. In section~\ref{s:heterogeneity}, I therefore extend the model to allow for heterogeneous labor and revisit the results.

\subsection{Impact on local workers}

I now determine the impact of in-migration on local workers using figure~\ref{f:inmigration}. Before migrants join the local labor market, the labor force and labor supply are given by dashed lines. After migrant workers join the labor force, the labor force and labor supply shift outward. The labor force and labor supply after in-migration are given by solid lines. The labor demand is unaffected by migration here, as wages are assumed to be fixed. The pre-migration solution of the model is at the intersection of the labor demand curve and old, dashed labor supply curve. The post-migration solution of the model is at the intersection of the labor demand curve and new, solid labor supply curve.

Figure~\ref{f:inmigration} shows that after in-migration, more local workers are unemployed, and fewer are employed. The old, dashed labor supply remains the local labor supply: the number of local workers who are employed as a function of labor market tightness. As migrant workers enter the labor market, tightness drops, and fewer local workers are employed: a movement down the old labor supply curve marked by the arrow in the figure. The local workers who are not employed any more move into unemployment.

The reason why tightness drops is simple. The labor demand curve is downward sloping, so the number of jobs in the labor market is somewhat limited. This means that firms are not able to absorb all the migrant workers arriving into the labor market. As there are more applicants for each job, local workers take longer to find jobs: they are more likely to be unemployed and less likely to be employed. In that way, the incoming workers take some jobs away from local workers. 

However, the aggregate number of jobs is not fixed. With in-migration, the labor market moves down the labor demand curve, so while local employment declines, aggregate employment increases.

The following proposition describes these results formally:
\begin{proposition}
The elasticity of labor market tightness with respect to migration is:
\begin{equation}
\e^{\t}_{I} = - \frac{1}{(1-\h) u(\t) + \frac{1-\a}{\a}\cdot \h\tau(\t)}. 
\label{e:thetai}\end{equation}
The elasticity is negative, so in-migration leads to a decrease in labor market tightness. Such a decrease causes an increase in local unemployment and a decrease in local employment. It also causes an increase in aggregate employment.
\label{p:migration}\end{proposition}

The proof of the proposition is relegated to appendix~\ref{a:proofs}. Expression \eqref{e:thetai} confirms that the results require diminishing returns to labor ($\a>0$). Indeed, when $\a\to 0$, the denominator becomes infinite, so $\e^{\t}_{I} \to 0$: labor market tightness does not respond to migration, so local unemployment is unaffected. In figure~\ref{f:inmigration}, with constant returns to labor instead ($\a=0$), the labor demand curve would be horizontal, so the outward shift in labor supply caused by in-migration has no effect on tightness. Thus, local workers find jobs at the same pace, and local unemployment is unchanged. This is what happens in the DMP model (appendix~\ref{a:dmp}). 

In the model, after a wave of in-migration, the job-separation rate of local workers is unaffected: local workers do not lose their jobs at a higher rate. But, once local workers lose their jobs, they remain unemployed longer since their job-finding rate is lower---as they compete with newly arrived workers. This mechanism is consistent with evidence from Germany. \citet[p.~438]{DSS17} find that the increase in unemployment for German workers when Czech commuters arrived was caused by reduced inflows into employment, not increased outflows from employment: German workers did not lose their jobs at a higher pace, but it became harder for those who did not have a job to find one. The mechanism also reconciles the evidence presented in section~\ref{s:evidence}, that in-migration induces local unemployment, and the evidence presented by \citet{WZ99}, that local workers are not laid off at a higher rate to be replaced by immigrant workers.

Public anxieties about migration often focus on the idea that migrant workers ``steal jobs'' from local workers. For instance, \citet[table~C3]{HWZ17} report that Austrian voters worry about the negative effects of immigration on the labor market: they think that immigrants threaten their labor market opportunities. Given the results from proposition~\ref{p:migration}, these anxieties are unsurprising. First, the job-finding rate for local jobseekers falls when immigrants arrive, so it becomes harder to find a job. Locals have fewer jobs available to them because the number of available jobs did not scale up with the increase in labor-force participants. Second, the unemployment rate of local workers increases when immigrants arrive. So local workers might feel that immigrants steal their jobs: the fraction of local workers who have a job is indeed lower, as the fraction who remain unemployed is higher. And the reason is that immigrant workers are now employed in some of the available jobs, relegating local workers to unemployment.

\section{In-migration when wages respond}\label{s:wage}

For simplicity, section \ref{s:inmigration} assumes that wages are fixed and that only labor market tightness responds to migration. In general, of course, both tightness and wages might respond to migration. This section generalizes the analysis so wages also respond to migration. 

\subsection{Modeling the response of wages to migration}

To capture the effect of migration on wages, I assume that the real wage is not constant but is a function of the migration factor:
\begin{equation}
w = \o \cdot I^{-\b},
\label{e:wage}\end{equation}
where $\o>0$ governs the wage level and $\b\in [0,\a)$ captures the flexibility of wages with respect to migration. If $\b=0$, wages do not respond to migration at all. If $\b>0$, wages fall when new workers enter the labor force because of wage competition.

Under this wage norm, the migration factor enters the labor demand as follows:
\begin{equation}
L^d(\t,I) =\bc{\frac{(1-\a) a I^{\b}}{\bs{1+\tau(\t)}^{1-\a}\o}}^{\frac{1}{\a}}.
\label{e:ldi}\end{equation}
When $\b>0$, in-migration stimulates labor demand by reducing wages.

\subsection{Impact on local workers}

What is the impact of in-migration when wages respond to migration? I follow the same steps as in section~\ref{s:inmigration} but allow for a wage response, which affects the labor demand as shown in \eqref{e:ldi}.

Figure~\ref{f:wage} illustrates the effects of in-migration. As in figure~\ref{f:inmigration}, where wages are fixed, the labor force and labor supply shift outward from the old, dashed lines to the new, solid lines when new workers enter the labor market. Because wages fall somewhat, the labor demand rotates outward from the old, dashed line to the new, solid line. As the labor demand is boosted by lower wages, tightness is lower after the influx of migrant workers, but not as low as when wages are fixed. Since tightness is lower, fewer local workers are employed: a movement down the old labor supply curve marked by the arrow in the figure. Hence, the number of unemployed local workers and the local unemployment rate increase, but not as much as when wages are fixed.

Collecting these results yields the following proposition:
\begin{proposition}
The elasticity of labor market tightness with respect to migration is:
\begin{equation}
\e^{\t}_{I} = -\bp{1-\frac{\b}{\a}} \cdot \frac{1}{(1-\h) u(\t) + \frac{1-\a}{\a}\cdot \h\tau(\t)}. 
\label{e:thetaiwage}\end{equation}
The elasticity is negative, so even when wages fall after a wave of in-migration, in-migration leads to a decrease in labor market tightness. This decrease causes an increase in local unemployment and a decrease in local employment. However, the effects are muted by a factor $1-\b/\a\in (0,1)$ compared to when wages are fixed. In-migration also causes an increase in aggregate employment.
\label{p:wage}\end{proposition}

The proof of the proposition is in appendix~\ref{a:proofs} as well. The intuition for the results remains the same, except that here, it is the combined drops in tightness and wages that absorb the influx of migrants. This means that both employed and unemployed local workers are hurt by in-migration. For unemployed workers, it is harder to find a job; for employed workers, labor income is reduced. However, the reduction in job-finding rate experienced by jobseekers is less severe than when wages do not respond. In that way, the drop in wages spreads the negative impact of in-migration across all workers---employed and unemployed.

The elasticity \eqref{e:thetaiwage} shows that the results require somewhat rigid wages ($\b<\a$). With flexible wages instead ($\b=\a$), the elasticity of tightness with respect to migration is 0, so tightness does not respond to migration. As tightness remains the same, local workers would find jobs at the same pace, and nothing happens to local unemployment and employment. In figure~\ref{f:wage}, with flexible wages, wages would fall more after a wave of in-migration, so the labor demand curve would respond more. The labor demand curve would shift in tandem with the labor supply curve, so the outward shifts in the two curves would have exactly offsetting effects on tightness. In that case, all adjustments to in-migration occur through wages; tightness does not respond.

\subsection{The local employment-wage disconnect}

The matching model breaks the usual association between local wage and employment responses that appears in the Walrasian world.\footnote{See for instance figure 1 in \citet{RR07}.} In a Walrasian model, local employment and wages either both fall after in-migration---with a downward-sloping labor demand curve---or both do not respond to migration---with a horizontal labor demand curve. Furthermore, the responses move in tandem, as determined by the slope of labor demand: the drops in local wages and local employment are both larger whenever the labor demand curve is steeper.

In the matching model, this association breaks down. If wages fall little with in-migration, then labor demand is barely boosted by in-migration, so tightness falls significantly with in-migration, which means that local unemployment rises significantly and local employment declines significantly (figure~\ref{f:inmigration}). On the contrary, if wages fall significantly with in-migration, then labor demand is boosted by in-migration, so tightness falls less, which implies that local unemployment rises little and local employment does not decline much (figure~\ref{f:wage}).

Through this mechanism, the model predicts either a weak wage drop and large drop in local employment, or a strong wage drop with a small drop in local employment. This employment-wage disconnect seems consistent with a lot of evidence. \citet[p.~175]{G12} finds for instance ``a displacement effect of 3.1 unemployed workers for every 10 immigrants that find a job, but no effect on relative wages.'' Similarly, \citet[p.~435]{DSS17} find ``a moderate decline in local native wages and a sharp decline in local native employment.'' 

In France, certain contracts provide much stronger wage protection than others. As predicted by the model, workers whose wages are protected see a significant increase in unemployment rate when immigrants enter the labor market: the arrival of 100 immigrants pushes 30--50 workers into unemployment \citep[p.~2]{E16}. For workers whose wages are not protected, however, the impact of immigration is different: wages fall significantly but the unemployment rate is unaffected \citep[p.~3]{E16}. Because most workers have contracts that protect their wages, in aggregate it is tightness and not wages that responds to immigration. Similarly, \citet{BE21} find that in France the labor market for female workers exhibited no wage response but strong tightness response to immigration, while the labor market for male workers operated the other way around---no tightness response but sharp wage response. 

\subsection{Response of aggregate demand to migration}\label{s:ad}

The model focuses on the impact of migration on the labor supply, but migrants who enter a local economy also affect aggregate demand, since they consume goods and services where they live. Fortunately, the aggregate-demand response to migration is incorporated through the response of real wages. 

When new workers arrive into the local economy, aggregate demand is boosted, which may raise the prices of locally provided goods and services. Most wages are nominally rigid, and because of it, changes in aggregate demand affect labor demand. As aggregate demand increases and prices rise, nominally rigid wages rise less than prices, leading to a drop in real wages, which stimulates labor demand.\footnote{\citet{GH84} discuss this mechanism. \citet{AS10} and \citet{MV14} find that prices of local goods went up when refugees from Burundi and Rwanda arrived in Tanzania in the 1990s. The fact that local prices rise after in-migration is itself evidence that this channel cannot overturn the results: it implies that the aggregate supply does not scale up in proportion to the aggregate demand, which itself suggests that labor demand does not scale up with aggregate demand, so that jobs are somewhat limited. If aggregate supply scaled up in proportion to aggregate demand, prices would not increase. The aggregate-demand response therefore only attenuates migration-induced unemployment, through a larger $\b$, but it does not eliminate it.}

Since the model allows real wages to fall with migration, as displayed in \eqref{e:wage}, the model actually incorporates the response of aggregate demand to migration. It suffices to interpret and calibrate the elasticity $\b$ as the response of real wages to migration generated both by lower nominal wages due to higher labor supply and by higher prices due to higher aggregate demand.\footnote{\citet{IZ19} find that this mechanism might be at play when immigrants enter Germany: the arrival of immigrants boosts the demand for services, which leads to higher service prices relative to service wages, which in turn stimulates job creation in the service sector.}

\section{Impact of in-migration on firms}\label{s:firms}

Because the model behaves similarly under a fixed wage and under the general wage norm \eqref{e:wage}, I will assume from now on that wages are given by \eqref{e:wage}. So all the results in the sections below hold whether wages are fixed or merely somewhat rigid.

A direct implication of the fact that labor market tightness declines with in-migration (see propositions~\ref{p:migration} and~\ref{p:wage}) is that firms benefit from in-migration, due to easier recruitment:
\begin{corollary}
In-migration leads to a reduction in the recruiter-producer ratio, an increase in the aggregate number of producers, and an increase in real profits. These results hold whether wages respond to migration or not.
\label{c:firms}\end{corollary}

A little bit of algebra is required to see the impact of in-migration on real profits. Real profits are given by 
\begin{equation*}
\pi = a P^{1-\a} - [1+\tau(\t)] w P,
\end{equation*}
where the first term is the firm's output and the second term is the real wage bill. The first-order condition of the firm problem \eqref{e:foc} links the real wage $w$ to the number of producers $P$: 
\begin{equation*}
w = \frac{(1-\a) a P^{-\a}}{1+\tau(\t)}.
\end{equation*}
Therefore the real wage bill is
\begin{equation*}
[1+\tau(\t)] w P = (1-\a) a P^{1-\a}.
\end{equation*}
Combining these results yields real profits as a function of the number of producers: 
\begin{equation}
\pi = \a \cdot a \cdot P^{1-\a}.
\label{e:profits}\end{equation}
This equation simply says that the economy's profit share $\pi/y$ is $\a$. Profits are therefore proportional to output. Since output is divided between labor income and profits, the economy's labor share is $1-\a$.

After in-migration, aggregate employment $L$ goes up, and tightness $\t$ falls, so the recruiter-producer ratio $\tau(\t)$ falls. Hence, the number of producers $P = L/[1+\tau(\t)]$ goes up. Therefore, equation~\eqref{e:profits} shows that profits go up after a wave of in-migration.

The model reveals winners and losers from migration. Firm profits increase after in-migration, so firm owners benefit from in-migration (corollary~\ref{c:firms}). Local workers, facing a higher unemployment rate and potentially lower wages, experience a decline in labor income from in-migration (proposition~\ref{p:wage}). Thus, the benefits from in-migration accrue to capitalists while the costs are borne by local workers.\footnote{\citet{AM26} reach a similar conclusion about the incidence of migration through a different channel. In their model firms exercise more monopsony power over migrants than over natives and cannot perfectly wage discriminate, so in-migration raises wage markdowns. They find that this expands aggregate native income while redistributing it from workers to firms---just like in this model.} If workers own a share of firms, the impact of in-migration is murkier: in-migration reduces their labor income but raises their capital income. If workers own very little capital, however, they are bound to be negatively affected by in-migration.

\section{In-migration in good and bad times}\label{s:businesscycle}

I now contrast the effects of migration on the labor market in good times---when jobs are abundant and tightness is high---and bad times---when jobs are scarce and tightness is low. The model reveals that the impact of in-migration on unemployment is not uniform across all economic conditions, but it is more severe in bad times than in good times.

\subsection{Modeling good and bad times}

In the United States, labor market fluctuations are driven by labor demand shocks, not labor supply shocks \citep{MS15}. I therefore model good and bad times as periods with high and low labor demand. Labor demand is governed by the wage-to-productivity ratio, $w/a$ (equation~\eqref{e:ld}). Good times are periods when the wage-to-productivity ratio is low so hiring workers is particularly profitable and the labor demand is elevated (figure~\ref{f:goodtimes}). Bad times are periods when the wage-to-productivity ratio is high so hiring workers is not very profitable and the labor demand is depressed (figure~\ref{f:badtimes}). 

What causes changes in the wage-to-productivity ratio, $w/a$? These fluctuations are typically driven by productivity shocks under a fixed wage $w$ \citep{H05}, or under a partially rigid wage $w = \o \cdot a^{1-\g}$, where $\g>0$ measures the rigidity of wages with respect to productivity \citep{BG10}.

In this section, I assume a wage norm that is as general as possible:
\begin{equation}
w = \o \cdot a^{1-\g} \cdot I^{-\b},
\label{e:wageGeneral}\end{equation}
where $\g\in(0,1]$ governs the rigidity of real wages with respect to productivity and $\b\in[0,\a)$ governs the flexibility of real wages with respect to migration. 

Then the wage-to-productivity ratio becomes $w/a = \o \cdot a^{-\g} \cdot I^{-\b}$,
which is strictly decreasing in productivity and enters the labor demand as follows:
\begin{equation}
L^d(\t,I,a) =\bc{\frac{(1-\a) a^{\g} I^{\b}}{\bs{1+\tau(\t)}^{1-\a}\o}}^{\frac{1}{\a}}.
\label{e:lda}\end{equation}
The demand equation shows that labor demand is increasing in productivity $a$ (since $\g>0$). So good times are represented by a high labor productivity and bad times by a low labor productivity.

\subsection{Labor market in good and bad times}

Figure~\ref{f:shock} allows me to obtain a range of comparative statics describing the business cycle. The figure shows that when labor productivity decreases, the labor demand curve shifts inward. As a result, labor market tightness drops, as the labor market moves down the labor supply curve. The drop in tightness has a range of implications, since all the other labor market quantities are functions of tightness.

The following proposition formalizes the results:
\begin{proposition}
The elasticity of tightness with respect to productivity is:
\begin{equation}
\e^{\t}_a = \frac{\g}{\a} \cdot \frac{1}{(1-\h) u(\t) + \frac{1-\a}{\a}\cdot \h\tau(\t)}.
\label{e:thetaa}\end{equation}
The elasticity is positive, so when labor productivity decreases, labor market tightness falls. Therefore, when productivity decreases, job seeking becomes harder: the job-finding rate falls and the unemployment rate increases. At the same time, recruiting becomes easier: the job-filling rate increases and the recruiter-producer ratio falls.
\label{p:businesscycle}\end{proposition}

Appendix~\ref{a:proofs} provides a formal proof. The key assumption that produces labor market fluctuations in response to productivity shocks is wage rigidity ($\g>0$): when wages are more rigid (higher $\g$), the response of tightness and the other variables is larger; if wages were flexible ($\g=0$), tightness and the other variables would not respond at all. In figure~\ref{f:shock}, when wages are flexible, the labor demand curve simply does not respond to productivity shocks, so these shocks are neutral.

\subsection{Impact of in-migration in good and bad times}

I now examine how the amount of unemployment induced by migration varies over the business cycle. A wave of in-migration increases local unemployment, so the model produces the type of job stealing that local workers commonly complain about. In addition, such job stealing is worse in bad times, as can be seen by comparing figure~\ref{f:goodtimes} to figure~\ref{f:badtimes}. The source of the results is that the labor supply curve is sharply convex. In good times, the labor supply curve is steep relative to the labor demand curve, so when it shifts out, it raises aggregate employment almost one-for-one. As a result, many of the newcomers are hired in new jobs. In bad times, the labor supply curve is flat relative to the labor demand curve, so when it shifts out, aggregate employment barely changes. In that case, the newcomers who are hired take jobs that previously went to local workers.

The following proposition formalizes the results:
\begin{proposition}
The elasticity of local employment with respect to migration is
\begin{equation}
\e^{N}_{I}(\t) = - \bp{1-\frac{\b}{\a}} \cdot \frac{1}{1+\frac{1-\a}{\a} \cdot \frac{\h}{1-\h}\cdot \frac{\tau(\t)}{u(\t)}}. 
\label{e:elh2}\end{equation}
The elasticity is negative, and its amplitude is decreasing with tightness. As a result, when labor market conditions deteriorate, the elasticity becomes more negative. The elasticity tends to $0$ when tightness tends to $\bar{\t}$, it equals $\b-\a<0$ when tightness is efficient, and it tends to $\b/\a - 1 < \b - \a$ when tightness goes to $0$. If wages do not respond to migration ($\b=0$), the elasticity is $-\a$ when tightness is efficient, and it falls to $-1$ when tightness goes to $0$.
\label{p:stealing}\end{proposition}
	
The proof is relegated to appendix~\ref{a:proofs}. The proposition shows that job stealing is more prevalent in bad times. Formally, the percentage reduction in local employment due to a migration-induced one-percent increase in the labor force is larger in bad times, when tightness is low, than in good times, when tightness is high. Intuitively, when the labor market is more depressed, firms do not benefit much from the presence of migrant jobseekers. This is because there are already so many jobseekers available. So firms do not create many new jobs when migrant workers arrive in bad times; therefore, migrant workers end up taking more jobs away from local workers.

Because fewer jobs are available in bad times, in-migration reduces local employment more drastically in bad times. In line with this prediction, \citet{EO23} find that immigration has a negative effect on employment in the short run, and this negative effect is stronger in regions that are experiencing worse economic conditions.

The worst-case scenario occurs when the labor market is the slackest. Then, if wages do not respond to migration, a migration-induced one-percent increase in the labor force leads to a one-percent decrease in local employment. Indeed, when the labor market is extremely slack, job rationing is the most stringent, so the number of jobs is almost fixed. With a fixed number of jobs, local employment and migration factor are related by $N \cdot I= \text{constant}$ so the elasticity of local employment with respect to migration is $-1$.

In that way, the labor market becomes closer to zero-sum in bad times, perhaps explaining why the pushback against immigration intensifies in depressed labor markets. \citet{H74} shows for instance that high unemployment and cut-throat competition for existing jobs during the Great Depression intensified anti-Mexican sentiment, leading to the pressured or forced repatriation of half a million Mexicans from the United States between 1929 and 1939.

In bad times, local workers are more likely to be opposed to in-migration because it hurts their labor market prospects more. In line with this prediction, Austrian voters shift more toward the anti-immigration FPO in response to immigration when local unemployment is higher. The effect of immigration on voting is twice as large in local labor markets with unemployment rates in the top quartile than in those with unemployment rates in the bottom quartile \citep[table~5A]{HWZ17}.

Relatedly, the number of immigration-control audits conducted by the US government increases with local unemployment: 1 percentage point more of unemployment doubles the number of audits within a state \citep[p.~141]{MS14}. A possible reason is that legislators push for stricter enforcement of immigration laws when unemployment is high because they face more pressure from their constituents---who want to reduce what they see as job stealing \citep[p.~139]{MS14}.

\section{Optimal migration policy}\label{s:welfare}

Lastly, the model has implications for the design of migration policy. In the short run in-migration hurts local workers---especially in bad times---but helps local firms. To understand the overall impact of in-migration, I assess the effect of in-migration on total local welfare---the welfare of local workers plus the welfare of local firm owners. The model suggests that a procyclical in-migration policy is optimal. This policy involves encouraging in-migration during periods when labor markets are inefficiently tight and limiting it when labor markets are inefficiently slack. Such a policy balances the benefits of in-migration---the increased labor supply that makes it easier for firms to fill vacant jobs---with the costs of in-migration---the diversion of labor income from local workers to migrant workers, particularly during times of economic hardship.

\subsection{Computing local welfare}

In the model, there is only one consumption good, produced by firms, which goes to local workers through local labor income, to immigrant workers through their labor income, and to firm owners through local profits. The goal is to assess how in-migration impacts local welfare, which is determined by the sum of local labor income and local profits.

Aggregate real profits are given by $\pi = y(P) - [1+\tau(\t)] w P$. Output can be rewritten as a function of the marginal product of labor: $y(P) = y'(P)P/(1-\a)$. Moreover, along the labor demand curve, the marginal product of labor is related to the wage and recruiter-producer ratio: $y'(P) = [1+\tau(\t)] w$. Combining these results, I express real profits as a function of the wage bill $w L$:
\begin{equation}
\pi(\t) =\bs{\frac{1}{1-\a}-1} [1+\tau(\t)] w P = \frac{\a}{1-\a} w L.
\label{e:pi}\end{equation}
From \eqref{e:pi}, I express local profits as a function of migration rate $i$ and local employment $N = L/(1+i)$:
\begin{equation*}
\pi = \frac{\a (1+i)}{1-\a} w N.
\end{equation*}
The local labor income is just the real wage times local employment: $w N$.

For simplicity, I leave distributional considerations out of social welfare, so local welfare is completely determined by local income.\footnote{Distributional considerations can be excluded by assuming that workers and firm owners have the same linear utility function. They can also be excluded by assuming that workers have access to firm profits, so everyone consumes the same. Workers have access to profits if workers own all firms and therefore receive their profits, or if profits are fully taxed and redistributed to workers.} Adding local profits and local labor income gives local welfare:
\begin{equation}
\Nc = \frac{1+\a i}{1-\a} \cdot w N.
\label{e:welfare}\end{equation}

\subsection{Effect of infinitesimal in-migration on welfare}

As a first step, I assess whether any in-migration might ever improve welfare. The following proposition states the results:
\begin{proposition}
In any labor market that is inefficiently slack, allowing in-migration reduces local welfare. In any labor market that is inefficiently tight, allowing in-migration improves local welfare. When the labor market is efficient, in-migration has no first-order effect on local welfare. These results hold whether wages respond to migration or not.
\label{p:welfare}\end{proposition}

The proof is relegated to appendix~\ref{a:proofs}. The proposition shows that when the labor market is inefficiently slack, a further decrease in tightness---keeping labor force constant---reduces welfare. From the perspective of local workers, a drop in tightness caused by in-migration is equivalent to a drop in tightness keeping labor force constant. The welfare generated by the labor-force increase, keeping tightness constant, goes to migrants and does not count toward local welfare. Hence, in-migration reduces local welfare when the labor market is inefficiently slack by further reducing tightness. Conversely, in-migration improves local welfare when the labor market is inefficiently tight by bringing tightness toward efficiency.

The result that immigration improves welfare when the labor market is excessively tight explains why governments have historically relied on immigration to address labor shortages in their economies. \citet[section~5.1]{HWZ17} provide a first example of such a policy. In the postwar boom of the 1950s and 1960s, growing labor shortages were hindering the Austrian economy. To alleviate those shortages, the Austrian government enacted policies designed to attract foreign workers from southern Europe. More immigrant workers were allowed in regions of Austria where unemployment was particularly low. \citet{BRS21} provide another, more recent example. Between 1999 and 2003, Switzerland opened its labor market to foreigners because the country faced a low unemployment rate and high demand for skilled workers. As expected, firms benefited from the reform---especially those that had faced the largest difficulties in recruiting skilled workers.

In the model, local welfare is the sum of local labor income plus local firm profits, so local welfare is the same as total local income. The results of proposition~\ref{p:welfare} can therefore help explain the findings by \citet{CCI24} that unauthorized immigration boosts the financial health of local governments when the local labor market is tight, but it damages their financial health when the local labor market is slack. Since unauthorized immigrants do not pay local taxes, the taxes paid by local workers and firm owners must go up when the labor market is tight and down when the labor market is slack, which occurs naturally if local income goes up with immigration when the labor market is tight and down when the labor market is slack.

\subsection{Optimal in-migration over the business cycle}\label{s:optimal}

Proposition~\ref{p:welfare} shows that in-migration improves welfare when the labor market is too tight. I now turn to the next question: what is the optimal amount of in-migration when the labor market is initially too tight? The proposition below states the results:
\begin{proposition}
From an initial situation with no migration and an inefficiently tight labor market, the optimal amount of in-migration brings the labor market to an inefficiently slack situation. From an initial situation with no migration and an inefficiently slack labor market, the optimal amount of in-migration is zero.
\label{p:optimal}\end{proposition}

The proof is in appendix~\ref{a:proofs} as well. The most important insight from the proposition is that if migration is the only tool available to policymakers, and if policymakers aim to maximize local welfare, then the labor market should always be inefficiently slack. Otherwise migration is suboptimal: more in-migration would improve welfare.

What is the intuition for the result that optimal migration brings the labor market to a slack situation? At the efficient tightness, a drop in tightness reduces labor income, but this reduction is entirely offset by an increase in local profits, so local welfare (the sum of labor income and profits) is unaffected. As in-migration increases, the share of labor income going to local workers shrinks, while all profits continue to go to local firm owners. So as in-migration increases, the profit motive plays an increasingly large role in welfare. This means that at the efficient tightness $\t^*$, if in-migration is already positive, a decrease in tightness raises local profits more than it reduces labor income---which means that a drop in tightness raises welfare. Accordingly, it is optimal to allow some more in-migration to reduce tightness further, into the slack territory.

\subsection{Feasibility of a cyclical migration policy}

Given the documented effects of migration on unemployment, a cyclical migration policy would be desirable. Such a policy might not be difficult to implement. In fact, in 2008 France enacted a policy that makes it easier to hire immigrant workers for firms facing a high labor market tightness \citep{S24}. The policy was targeted to narrow skills, but it is conceivable that a similar policy could be implemented based on the national labor market tightness.

In the United States, a cyclical immigration policy could be implemented just like unemployment insurance.\footnote{The duration of unemployment insurance is automatically extended from 26 weeks to 39 weeks when the unemployment rate in a state reaches 6.5\% and then to 46 weeks when the state unemployment rate reaches 8\% \citep{M17}.} The number of immigrants authorized in the country or in a state could depend on the national or state unemployment rate. More immigrants could be allowed to enter when the labor market is inefficiently tight, alleviating recruiting difficulties for firms, and fewer when the labor market is inefficiently slack, protecting local workers who are already struggling to find jobs.

In practice, it is difficult to forecast business cycles. If the economy is hot and many migrant workers are brought in to cool the labor market, but the economy unexpectedly turns around, the policy might not improve welfare. Fixed-term work visas might alleviate this issue. In the postwar period, 1945--2020, the average US business cycle lasts 6 years \citep{NBER23}. So multi-year work visas would squarely fit within the business-cycle timeframe. 

\section{Political support and opposition to immigration}\label{s:politics}

This section turns to the political implications of the model. The model predicts the short-run impact of migration on two constituencies: workers and firm owners. From this, I infer the political support that immigration might receive from different groups under different circumstances.

\subsection{Opposition to immigration from pro-labor parties} 

Pro-labor parties design their programs to improve the welfare of workers. Since immigration hurts the welfare of local workers, such parties are expected to oppose immigration. Moreover, since the elasticity of local employment with respect to migration is more negative in bad times, opposition to immigration should be louder in bad times. For the same reasons, labor unions are expected to be opposed to immigration, since immigration reduces labor income. 

Evidence from the United States supports this prediction. The American Federation of Labor (AFL) and other white working-class activists were pushing for restrictions on immigration at the end of the 19th century and beginning of the 20th century \citep{F64,J21}. 

Lobbying around the Chinese Exclusion Act provides a good illustration. The Chinese Exclusion Act was passed in 1882 to block immigration from Chinese laborers into the United States. The Act initially lasted 10 years, was extended for another 10 years in 1892, and then made permanent in 1902 \citep[p.~8]{LMQ24}. In 1881, before the Act was enacted, the AFL adopted the following resolution at their convention: ``Whereas the experience of the last thirty years in California and on the Pacific coast having proved conclusively that the presence of Chinese and their competition with free white labor is one of the greatest evils with which any country can be afflicted: Therefore be it resolved, that we use our best efforts to get rid of this monstrous evil'' \citep[p.~25]{AFL02}. In 1902, the leaders of the AFL told Congress that letting the Act lapse would result in ``hundreds of thousands of our citizens to be deprived of employment to make room for this Asiatic coolie'' \citep[p.~23]{AFL02}. They also wrote that before the Act, white workers were losing jobs in large numbers: ``Soon after the negotiation of the Burlingame treaty in 1868 large numbers of Chinese coolies were brought to this country under contract. Their numbers so increased that in 1878 the people of [California] made a practically unanimous demand for the restriction of immigration. Our white population suffered in every department of labor and trade, having in numerous instances been driven out of employment by the competition of the Chinese'' \citep[p.~25]{AFL02}.

Similarly, \citet[p.~26]{H74} explains that the repatriation of Mexican immigrants from the United States during the Great Depression was championed by ``small farmers, progressives, labor unions, eugenicists, and racists.'' And after World War 2, German unions also opposed the entry of foreign workers by arguing that they competed for jobs with native workers \citep[p.~24]{AT24}.

\subsection{Support for immigration from pro-business parties}

Pro-business parties, by contrast, design their programs to support business and firm owners. Since immigration improves the profits of firms, such parties are expected to favor immigration.

The debate around the Chinese Exclusion Act again provides a good illustration. While labor organizations supported the Act, business owners generally opposed it \citep{LMQ24}. These business owners were supported by pro-business Republicans, who welcomed Chinese workers \citep[p.~8]{J21}. One such businessman who testified in Congress on behalf of Chinese workers is Charles Crocker, who was president of the Southern Pacific Railroad, and who hired a large number of Chinese immigrants for the construction of the first transcontinental railroad. In Congress, he argued that ``without Chinese labor we
would be thrown back in all the branches of industry, farming, mining, reclaiming lands, and everything else'' \citep[p.~19]{Z19}. 

Similarly, Mexican immigration into the United States in the 1920s and 1930s was supported by ``large-scale growers of sugar beets, cotton, and vegetables, allied with railroads, chambers of commerce, and business associations'' \citep[p.~26]{H74}. 

\section{Numerical illustration}\label{s:illustration}

This section illustrates numerically the results of the previous sections.

\subsection{Calibration of the model}\label{s:calibration}

I first calibrate the model based on US evidence. A critical parameter in the context of migration is the elasticity of real wages to the labor force, $\b$. This parameter determines how much wages change with migration and thus determines the amplitude of the results. Surveys of the literature conclude that immigration only generates modest effects on local wages. For instance, at the end of their survey, \citet[p.~839]{BK15} write:
\begin{quote}
While some studies have found important effects, overall it seems to us that most research does not find quantitatively important effects of immigration on native wage levels or the wage distribution.
\end{quote}
\citet{RR07} and \citet{FH95} share the same sentiment. To capture these modest effects, I simply set the wage elasticity to $\b=0$---so wages do not respond to migration.

Of course, some studies do find that wages fall with immigration.\footnote{See for instance \citet{B03,B17}, \citet{BR12a}, \citet{DSS17}, \citet{M17a}, \citet{M20}, or \citet{D24a}.} For completeness, appendix~\ref{a:wage} reproduces the simulations for a range of positive values for the wage elasticity $\b$. As the theory already showed, all the results remain qualitatively the same---only amplitudes change.

The rest of the parameters are standard so their calibration is relegated to appendix~\ref{a:calibration}. The calibrated parameter values are summarized in table~\ref{t:calibration}.

\begin{table}[t]
\caption{Parameter values in simulations}
\begin{tabular*}{\textwidth}{@{\extracolsep{\fill}}llp{7cm}}\toprule
Value & Description & Source or target\\
\midrule
\multicolumn{3}{c}{A. Calibration from direct measurement}\\
$\g = 0.3$ & Rigidity of real wages & \citet{M12} \\
$\b = 0$ & Migration elasticity of real wages & \citet{FH95}, \citet{RR07}, \citet{BK15}\\
$\k = 1$ & Recruiting cost & \citet{MS24} \\
$\l = 0.097$ & Quarterly job-separation rate & \citet{MS21b}\\
\midrule
\multicolumn{3}{c}{B. Calibration to match target}\\
$\a = 0.35$ & Decreasing returns to labor & Labor share $=65\%$ \citep{ADK20} \\
$\h = 0.48$& Unemployment elasticity of matching & Beveridge elasticity $=1$ \citep{MS24}\\
$\m = 2.16$ & Quarterly matching efficacy & $u^* = 4.3\%$ \citep{MS24}\\
$\o = 0.64$ & Level of real wages & $u = u^*$ when $ a =1$ (normalization)\\
\bottomrule\end{tabular*}
\note{The parameter values described in the table are obtained in appendix~\ref{a:calibration}.}
\label{t:calibration}\end{table}

\subsection{Simulation of the effects of in-migration}

This section simulates the calibrated model to quantify the effects of migration on unemployment at different stages of the business cycle (figure~\ref{f:illustration}).

\begin{figure}[p]
\subcaptionbox{Labor market tightness over the cycle\label{f:atheta}}{\normalgraphic[7]{\pdf}}\hfill
\subcaptionbox{Unemployment over the cycle\label{f:thetau}}{\normalgraphic[8]{\pdf}}\fspace
\subcaptionbox{Migration elasticity of local employment\label{f:thetaelh}}{\normalgraphic[9]{\pdf}}\hfill
\subcaptionbox{Migration-induced unemployment\label{f:thetastealing}}{\normalgraphic[10]{\pdf}}\fspace
\subcaptionbox{Migration elasticity of welfare\label{f:ewh}}{\normalgraphic[11]{\pdf}}\hfill
\subcaptionbox{Optimal in-migration over the cycle\label{f:istar}}{\normalgraphic[12]{\pdf}}
\caption{Simulation of the model over the business cycle and the effects of in-migration}
\note{The figure is obtained by simulating the labor market model under the calibration presented in table~\ref{t:calibration}. The pink lines represent an efficient labor market. Panel A plots $\t(a)$, computed by solving \eqref{e:solution}. Panel B plots $u(\t)$, computed using \eqref{e:u}. Panel C plots $\e^{N}_{I}(\t)$, computed using \eqref{e:elh2}. Panel D plots the number of local workers who become unemployed per 100 arrivals, given by $- 100 \times (N/H) \times \e^{N}_{I}(\t)$. Panel E plots $\e^{\Nc}_{I}(\t)$, computed using \eqref{e:ewm}. Panel F plots the optimal migration rate $i^*(\t)$, computed by solving \eqref{e:istar}.}
\label{f:illustration}\end{figure}

In the model, business-cycle fluctuations are generated by productivity shocks (figure \ref{f:atheta}). When productivity is below 1, the labor market is inefficiently slack: too many workers are unemployed. When productivity is at 1, the labor market is efficient. And when productivity is above 1, the labor market is inefficiently tight: too many workers are allocated to recruiting.

The response of tightness to productivity in the model is comparable to that observed in the United States, confirming that the model is not subject to \citet{S05}'s puzzle. In the United States, the elasticity of tightness with respect to productivity is 8.6 \citep[p.~1741]{M12}. In the model, when productivity falls from 1 to 0.95, tightness drops from 1 to 0.5. So when productivity falls by 5\%, tightness drops by 50\%, which corresponds to an elasticity of $50\%/5\% = 10$. Hence, with wage rigidity, the model produces realistic fluctuations in tightness in response to productivity shocks.

When tightness fluctuates, it becomes harder or easier for jobseekers to find jobs, so the unemployment rate fluctuates (figure~\ref{f:thetau}). When the labor market is efficient, tightness is 1 and the unemployment rate is 4.3\%. When tightness rises to 1.5, the unemployment rate falls to 3.5\%. When tightness falls to 0.5, the unemployment rate rises to 6.1\%. If tightness falls further to 0.25, unemployment rises further to 8.5\%.

From proposition~\ref{p:stealing} and the calibration in table~\ref{t:calibration}, I infer that the elasticity of local employment with respect to migration is $-1$ when tightness is $0$, it grows to $-\a = -0.35$ when tightness reaches 1, and it asymptotes to $0$ when tightness converges to $\bar{\t}$ (which is a very large number). The model simulations are useful to display how rapidly the migration elasticity of local employment falls when tightness drops (figure~\ref{f:thetaelh}). When tightness is 0.5, the elasticity falls to $-0.52$, implying that a wave of in-migration amounting to 1\% of the labor force reduces the number of local employees by 0.52\%. By contrast, when tightness is 1.5, the elasticity is only half of that, at $-0.26$, implying that the reduction in the number of local employees is only half of what it is with a tightness of 0.5. Hence, the response of local employment to migration is strongly state-dependent.

Accordingly, the amount of migration-induced unemployment is substantial and larger when the labor market is slacker (figure~\ref{f:thetastealing}). When labor market tightness is at its efficient level of 1, 33 local workers become unemployed when 100 migrants arrive in their labor market. When labor market tightness is 0.5, corresponding to a somewhat slack labor market, the number of local workers who become unemployed rises to 49. When labor market tightness is 0.1, corresponding to an extremely slack labor market, the number of local workers who become unemployed rises further to 73. By contrast, when labor market tightness is 2, corresponding to a very tight labor market, the number of local workers who become unemployed falls to 20.\footnote{When tightness becomes extremely low, below 0.05, the amount of migration-induced unemployment starts falling and converges to 0 when tightness goes to 0. This is just a mechanical effect: when tightness goes to 0, all workers become unemployed, so it becomes physically impossible for migration to push more employed workers into unemployment---since there are no employed workers left. This mechanical effect explains why the curve in figure~\ref{f:thetastealing} is not monotonic, whereas the elasticity $\e^{N}_{I}$ in figure~\ref{f:thetaelh} is monotonic.}

\subsection{Comparing simulation and experimental results}

The amount of migration-induced unemployment in the calibrated model is commensurate to the amount observed in migration experiments (table~\ref{t:evidence}). On average across experiments, 37 local workers become unemployed for 100 arrivals. In the model, when the labor market is efficient, 33 local workers become unemployed for 100 arrivals. In the United States, the labor market is generally inefficiently slack: tightness averages $0.73<1$ between 1930 and 2024 \citep[section~3D]{MS24}. At that tightness of 0.73, 40 local workers become unemployed for 100 arrivals. These simulated numbers are close to the experimental numbers. 

The range of results in the simulations is also comparable to the range of empirical results. \citet{H92} finds that the return of 100 repatriates from Algeria pushed 20 French workers into unemployment. This number might seem low, but the labor market was also extremely tight in France at the time. When repatriation started in 1962, the unemployment rate in France was only 1\% \citep[table~1]{H92}. At a tightness of 2, which is extremely high for the United States (since World War 2, it has only been reached during the pandemic), the calibrated model predicts that 20 local workers are pushed into unemployment for 100 arrivals. So the very tight labor market in France in the 1960s might explain why the repatriation only had muted effects on local employment. 

The same observations apply to the Portuguese returnees. \citet{M17a} finds that only 22 local workers become unemployed for 100 returns. Once again, this low number can be explained by the very tight labor market in Portugal at the time: the unemployment rate was only about 2\% in 1974 \citep[figure~4]{M17a}. Such a tight market helped absorb the returnees into the workforce.

By contrast, the European labor market was quite slack in the 1990s. In 1995, the unemployment rate in European OECD countries averaged 10.3\% \citep[table~1]{LS98}. Such slack might explain why \citet{AK03} found that immigration from Yugoslavia affected local workers so adversely in the 1990s. Their midpoint estimate is that 59 European workers were pushed into unemployment when 100 Yugoslavian workers entered their local labor market. This is comparable to the effect that in-migration has in the model for a tightness of 0.3, or equivalently an unemployment rate of 7.8\%.

\subsection{Simulation of the optimal migration policy}

Next, I simulate the calibrated model to quantify the optimal migration policy.

As predicted by proposition~\ref{p:welfare}, the elasticity of social welfare with respect to migration depends on the health of the labor market (figure~\ref{f:ewh}). This means that the marginal effect of migration on welfare depends on the health of the labor market. If the labor market is operating efficiently, migration has no marginal effect on welfare. If the labor market is inefficiently slack, a small wave of in-migration reduces welfare. By contrast, if the labor market is inefficiently tight, a small wave of in-migration improves welfare.

The optimal in-migration policy is sharply procyclical (figure~\ref{f:istar}). When the labor market is inefficiently slack, in-migration reduces welfare, so it is optimal to disallow it. When the labor market is inefficiently tight, in-migration improves welfare, so it is optimal to allow it. For instance, if the initial tightness is 1.2 instead of 1, then optimal in-migration is 1\% of the labor force. Such optimal in-migration would amount to 1.7 million immigrants in the United States today \citep{CLF16OV}. This is about twice the yearly amount of net in-migration to the United States over 2010--2019, and about half the yearly amount in 2023 \citep[p.~5]{CBO24}. More in-migration should be allowed when the labor market is initially tighter. If the initial tightness is 1.4 instead of 1.2, optimal in-migration doubles to 2\% of the labor force.

The simulations also show that when the labor market is initially too hot and in-migration is optimal, labor market tightness falls below 1, but not by much. The optimal tightness, which prevails under optimal migration, is 0.99 when the initial tightness is 1.2, and 0.98 when the initial tightness is 1.4. So roughly speaking, optimal in-migration should be sufficient to bring tightness down to its efficient level of 1.

\section{Heterogeneous labor}\label{s:heterogeneity}

This section extends the model to allow for heterogeneous labor. Just like in the model by \citet[section~2.3]{B99}, there are two types of labor: high skill and low skill.

\subsection{Modeling high-skill and low-skill labor}

Some migrant and local workers have more skill and some have less. Within each skill, locals and migrants remain perfect substitutes. This assumption is made for simplicity: it means that only the total numbers of high-skill and low-skill workers matter for the outcome of the model, and not the composition of each group. \citet[section~2.3]{B99} and \citet{A21} make the same assumption, and the evidence suggests that local and migrant workers with the same skill set might be close to perfect substitutes.\footnote{\citet{BGH08} report that they ``cannot reject the hypothesis that comparably skilled immigrant and native workers are perfect substitutes.'' \citet{OP12} estimate that while comparably skilled immigrant and native workers are not perfect substitutes, the elasticity of substitution is large, around 20.} 

The presence of two skill levels means that the labor market is now segmented into two markets: a low-skill market, indexed by $j = l$, and a high-skill market, indexed by $j = h$. There are $H_h$ workers in the high-skill labor market and $H_l$ workers in the low-skill labor market. Both labor markets have exactly the same structure as in the homogeneous-labor model, but with distinct matching functions, job-separation rates, recruiting costs, labor market tightnesses, and unemployment rates.

The representative firm produces output by combining $P_h$ high-skill producers and $P_l$ low-skill producers through a constant-elasticity-of-substitution (CES) aggregator:
\begin{equation}
x(P_h, P_l) = \bs{\z P_h^{(\s-1)/\s} + (1-\z) P_l^{(\s-1)/\s}}^{\s/(\s-1)},
\label{e:CES}
\end{equation}
with aggregation parameter $\z \in (0,1)$ and elasticity of substitution $\s > 1$. The restriction $\s>1$ means that low-skill and high-skill producers are gross substitutes; it will ensure that each skill market has a downward-sloping labor demand curve.\footnote{The condition $\s>1$ also implies that skill-biased technical change raises the skill premium; this is why theoretical models of the skill premium typically maintain it \citep[p.~1107]{AA11}.}

The firm's output then is
\begin{equation}
y(P_h, P_l) = a x(P_h, P_l)^{1-\a},
\label{e:Y}
\end{equation}
with productivity parameter $a > 0$ and diminishing-returns parameter $\a \in (0, 1)$.

The firm recruits workers in each skill market, in which tightness is $\t_{j}$ and the recruiting wedge is $\tau_{j}(\t_{j})$, where $\tau_{j}$ takes the form \eqref{e:tau}, and the unemployment rate is $u_{j}(\t_{j})$, where $u_{j}$ takes the form \eqref{e:u}. The number of employees of skill $j$ in the firm is $L_{j} = [1+\tau_{j}(\t_{j})] P_{j}$. For simplicity, real wages for each skill $j$ are fixed at $w_{j}$. Here, I focus on the case with fixed wages to highlight the response of the skill-specific tightnesses and unemployment rates to migration.

\subsection{Solving the firm problem}

The firm chooses the numbers of high-skill and low-skill producers $P_h$ and $P_l$ to maximize profit:
\begin{equation}
y(P_h, P_l) - \hat{w}_h(\t_h) P_h - \hat{w}_l(\t_l) P_l,
\label{e:firm}\end{equation}
where 
\begin{align}
\hat{w}_l(\t_l) &= [1+\tau_l(\t_l)] w_l\label{e:wl}\\
\hat{w}_h(\t_h) &= [1+\tau_h(\t_h)] w_h.\label{e:wh}
\end{align}
are the effective wages for producers of skill $l$ and $h$.

The problem can be solved in two stages, following the usual strategy \citep[pp.~139--141]{MWG95}. Let's define the scale of the firm operation by $X = x(P_h, P_l)$. First, given effective wages $(\hat{w}_h, \hat{w}_l)$ and scale $X$, the firm must choose the labor bundle $(P_h, P_l)$ to minimize input cost. The solution defines the conditional skill demands $P_h^c(\hat{w}_h, \hat{w}_l, X)$ and $P_l^c(\hat{w}_h, \hat{w}_l, X)$ and the cost function
\begin{equation*}
\hat{c}(\hat{w}_h, \hat{w}_l,X)=\min[P_h, P_l]{\hat{w}_h P_h + \hat{w}_l P_l\;\text{such that}\;x(P_h, P_l)\ge X}.
\end{equation*}
Second, the firm determines its scale of operation $X$ to maximize profits:
\begin{equation}
a X^{1-\a} - \hat{c}(\hat{w}_h, \hat{w}_l, X).
\label{e:twostep}\end{equation}

Since the production aggregator $x$ is homogeneous of degree 1, the cost function is linear in the scale of production:
\begin{equation}
\hat{c}(\hat{w}_h, \hat{w}_l, X) = c(\hat{w}_h, \hat{w}_l) \cdot X,
\end{equation}
where 
\begin{equation*}
c(\hat{w}_h, \hat{w}_l) = \min[P_h, P_l]{\hat{w}_h P_h + \hat{w}_l P_l\;\text{such that}\;x(P_h, P_l)\ge 1}
\end{equation*}
is the unit cost function. With CES aggregator~\eqref{e:CES}, the unit cost takes the usual form \citep{R95}:
\begin{equation}
c(\hat{w}_h, \hat{w}_l) = \bs{\z^{\s} \hat{w}_h^{1-\s} + (1-\z)^{\s} \hat{w}_l^{1-\s}}^{1/(1-\s)}.
\label{e:unitcost}\end{equation}
Shephard's lemma implies that the conditional skill demands equal the partial derivatives of the cost function with respect to the corresponding effective wage:
\begin{align}
P_l^c(\hat{w}_h, \hat{w}_l, X) &= \pd{c(\hat{w}_h,\hat{w}_l)}{\hat{w}_l}\cdot X\label{e:pld1}\\
P_h^c(\hat{w}_h, \hat{w}_l, X) &= \pd{c(\hat{w}_h,\hat{w}_l)}{\hat{w}_h}\cdot X \label{e:phd1}.
\end{align}
Combining the results with \eqref{e:unitcost} yields the following expressions for the conditional skill demands:
\begin{align}
P_l^c(\hat{w}_h, \hat{w}_l, X) &= (1-\z)^{\s} \hat{w}_l^{-\s} c(\hat{w}_h,\hat{w}_l)^{\s}\cdot X\label{e:pld2}\\
P_h^c(\hat{w}_h, \hat{w}_l, X) &= \z^{\s} \hat{w}_h^{-\s} c(\hat{w}_h,\hat{w}_l)^{\s}\cdot X\label{e:phd2}.
\end{align}

The second-step first-order condition is $(1-\a) a X^{-\a} = c(\hat{w}_h, \hat{w}_l)$, so the firm's optimal scale is
\begin{equation}
X^d(\hat{w}_h, \hat{w}_l) = \bs{\frac{(1-\a) a}{c(\hat{w}_h, \hat{w}_l)}}^{1/\a},
\label{e:xd}\end{equation}
where $c(\hat{w}_h, \hat{w}_l)$ is the unit cost \eqref{e:unitcost}.

Combining the scale demand with the conditional skill demands, and using $L_{j}=[1+\tau_{j}(\t_{j})] P_{j}$, I obtain the total demands for each skill $j$:
\begin{equation}
L_{j}^d(\t_l, \t_h) = [1 + \tau_{j}(\t_{j})] P_{j}^c(\hat{w}_h(\t_h), \hat{w}_l(\t_l), X^d(\hat{w}_h(\t_h), \hat{w}_l(\t_l))).
\label{e:ldj}\end{equation}
The demand for labor of skill $j$ expresses total skill-$j$ employment as a function of the two tightnesses. In the demand expressions, the conditional skill demands $P_{j}^c$ are given by ~\eqref{e:phd2} and \eqref{e:pld2}, the scale demand is given by \eqref{e:xd}, and the effective wages $\hat{w}_l$ and $\hat{w}_h$ are given by \eqref{e:wl} and \eqref{e:wh}.

\subsection{Solution of the model}

Supply of labor of skill $j$ retains the usual form, given by \eqref{e:ls}:
\begin{equation}
L_{j}^s(\t_{j}, H_{j}) = [1 - u_{j}(\t_{j})] H_{j}.
\label{e:Ls}\end{equation}
The solution of the heterogeneous-labor model is a pair of tightnesses $(\t_l, \t_h)$ such that supply equals demand in each skill market:
\begin{align}
L_l^d(\t_l, \t_h) &= L_l^s(\t_l, H_l) \label{e:equilibrium1}\\
L_h^d(\t_l, \t_h) &= L_h^s(\t_h, H_h) \label{e:equilibrium2}.
\end{align}
Just as with homogeneous labor, a solution always exists and is unique (appendix~\ref{a:solution}).

\subsection{Responses to in-migration}

I now consider a change in the low-skill labor force, $H_l$, which represents a low-skill in-migration shock. Just like in the simple case with homogeneous labor, I implicitly differentiate the supply-equals-demand equations \eqref{e:equilibrium1} and~\eqref{e:equilibrium2} to obtain the elasticities of tightnesses with respect to low-skill in-migration. I derive the following result, which describes what happens after an inflow of low-skill workers:
\begin{proposition}
An inflow of low-skill workers raises the low-skill unemployment rate. The effect on high-skill unemployment has the sign of $\s-1/\a$. If the skill groups are net substitutes ($\s>1/\a$): the substitution effect dominates the scale effect, so high-skill unemployment rises. If the skill groups are net complements ($\s<1/\a$): the scale effect dominates the substitution effect, so high-skill unemployment falls. In the knife-edge case $\s=1/\a$, the cross-skill effects vanish and the two skill markets are insulated.
\label{p:heterogeneity}\end{proposition}

The proof of the proposition is relegated to appendix~\ref{a:proofs}. The proposition shows that an inflow of low-skill workers always raises the low-skill unemployment rate by the within-skill mechanism inherited from the homogeneous-labor model. It also maps the long-standing empirical debate over substitutability between high-skill and low-skill labor onto a transparent threshold, and yields signed predictions for the cross-skill employment effects of migration on unemployment. If the skill groups are net substitutes, low-skill in-migration hurts high-skill workers by drawing firms away from high-skill producers. On the other hand, if the skill groups are net complements, more abundant low-skill labor expands output and pulls along high-skill demand, so low-skill in-migration actually helps high-skill workers.

Which case is the most realistic? For a labor share of about two-thirds, $\a = 0.35$, so the substitutability threshold is $1/\a \approx 2.9$. In their reviews of the literature, \citet[p.~1107]{AA11} find that $\s$ falls between 1.4 and 2.0 while \citet[p.~184]{OP12} place $\s$ between 1.5 and 2.5. So the overall tendency in the literature is that high-skill and low-skill workers are net complements: $\s < 1/\a$. Only a few estimates suggest that high-skill and low-skill workers are net substitutes. For example, \citet[p.~182]{OP12} propose a specification in which the elasticity of substitution is $\s = 3$. In that case, low-skill in-migration also hurts high-skill workers.

The same analysis would hold for a high-skill in-migration shock, in which $H_h$ rises, with the same results. High-skill in-migration always hurts high-skill workers but helps low-skill workers if and only if the skill groups are net complements.

\section{Conclusion}\label{s:conclusion}

The model in this paper offers a broader theory of migration, which captures the effects of migration not only on local wages but also on the local unemployment rate. Unlike in a Walrasian model, the labor market is organized around a matching function, so the adjustment to migration happens through both wages and labor market tightness, which itself determines workers' job-finding rate and the unemployment rate. After a wave of in-migration, more workers compete for jobs, so labor market tightness declines. As a result, local jobseekers take longer to find a job and face a higher unemployment rate.

By accounting for the response of tightness to migration, the model alleviates some of the theoretical tension between the Cardian and Borjasian branches of the immigration literature. The Borjasian branch supports a downward-sloping labor demand curve \citep{B03}. The Cardian branch believes that immigration does not reduce the wage of local workers \citep{C90}. In the Walrasian model, these two views are incompatible because the Cardian view requires a horizontal labor demand curve. In this paper's model, however, these two properties are compatible: the model accommodates a labor demand curve that is downward sloping and fixed wages. The crux is that the model contains an additional adjustment margin: it adjusts to migration not only through wages but also through labor market tightness. An immigration shock is an outward shift of the labor supply curve, which moves the labor market down the labor demand curve (figure~\ref{f:inmigration}). When wages are fixed, tightness absorbs the entire shock: it declines substantially. So immigration may adversely affect local workers even when it does not affect local wages---because it raises the local unemployment rate.

The paper's model can also be applied to the debate on immigration that followed the coronavirus pandemic in the United States. The US labor market was exceptionally tight in the recovery from the pandemic. The labor market became inefficiently tight in 2021 and labor market tightness reached 2.0 in 2022, a level it had not reached since the end of World War 2 \citep[figure~12]{MS24}. The Federal Reserve started raising interest rates in 2022, one year after the labor market turned inefficiently tight \citep[p.~374]{MS24}. However, the labor market was slow to cool: it only reached efficiency in April 2025, four entire years after turning inefficiently tight \citep{TIGHTNESS}. Therefore, allowing for some more immigration in 2021--2024 would have helped cool the labor market and might have enhanced social welfare. Immigration was substantial in 2022 and 2023 compared to previous years \citep[p.~5]{CBO24}. But it was not enough to bring labor market tightness below 1, so immigration remained below the model-implied social optimum.

\bibliography{\bib}

\appendix

\section{Impact of in-migration in DMP models}\label{a:dmp}

This appendix shows that in a \citet{D82}-\citet{M82}-\citet{P85} (DMP) model, there is some unemployment, but labor demand is perfectly elastic. Thus, migrant workers are absorbed into the labor market without affecting local workers.

The DMP model assumes constant instead of diminishing returns to labor ($\a=0$). With constant returns to labor, the labor demand is degenerate: firms' optimal employment choice solely determines tightness. Setting $\a=0$ in \eqref{e:ld} gives
\begin{equation}
\bs{1+\tau(\t)} w = a,
\label{e:ldstd}\end{equation}
which determines tightness in the model, independently from employment. This labor-demand relation holds irrespective of the wage-setting assumption. It holds with Nash bargained wages, as in the textbook DMP model \citep{P00}, or with fixed wages, as in the DMP model by \citet{H05}.

\begin{figure}[p]
\subcaptionbox{Unemployment-vacancy representation\label{f:dmpuv}}{\normalgraphic[13]{\pdf}}\hfill
\subcaptionbox{Employment-tightness representation\label{f:dmp}}{\normalgraphic[14]{\pdf}}\fspace
\subcaptionbox{In-migration if wages are fixed\label{f:dmpmigration}}{\normalgraphic[15]{\pdf}}\hfill
\subcaptionbox{In-migration if wages fall\label{f:dmpwage}}{\normalgraphic[16]{\pdf}}
\caption{Effects of in-migration in DMP models}
\note{The job-creation curve and labor demand curve (LD) represent the same locus of points, given by \eqref{e:ldstd}. The Beveridge curve and labor supply curve (LS) represent the same locus of points, given by \eqref{e:ls}. In panels C and D, the dashed lines represent the curves before migration, and the solid lines represent the curves after migration. In panel D, the arrow shows the increase in local employment after in-migration.}
\label{f:standard}\end{figure}

Here I analyze the DMP model in an employment-tightness diagram (figure~\ref{f:dmp}). It is more common to analyze the DMP model in an unemployment-vacancy diagram (figure~\ref{f:dmpuv}). However, the two presentations are equivalent. The labor demand curve in figure~\ref{f:dmp} and the job-creation curve in figure~\ref{f:dmpuv} both represent the tightness level defined by \eqref{e:ldstd}. The labor supply curve in figure~\ref{f:dmp} and the Beveridge curve in figure~\ref{f:dmpuv} both represent the locus of points satisfying \eqref{e:ls}. Equation \eqref{e:ls} links tightness $\t$ to the employment level $L$. But since $\t = v/u$ and $L = (1-u) H$, and since $f(\t) = \m\t^{1-\h}$, it is easy to rewrite it as a Beveridge curve linking the vacancy rate $v$ to the unemployment rate $u$:
\begin{equation*}
v(u) = \bp{\frac{\l}{\m}\cdot\frac{1-u}{u^{\h}}}^{1/(1-\h)}.
\end{equation*}
The Beveridge curve is decreasing and convex, as plotted in figure~\ref{f:dmpuv}. 

It is advantageous to study the DMP model in an employment-tightness diagram because the diagram offers a more interesting description of migration. In the unemployment-vacancy diagram, migration would not appear since it affects neither the job-creation curve nor the Beveridge curve. In the employment-tightness diagram, by contrast, a wave of in-migration shifts the labor supply curve (figure~\ref{f:dmpmigration}).

Let's turn to the effect of migration in the DMP model. The labor demand \eqref{e:ldstd} does not involve labor force $H$, so labor market tightness is the same irrespective of the amount of migration. This means that the unemployment rate is independent of labor force, and thus that migration has no effect on the unemployment rate. Figure~\ref{f:dmpmigration} illustrates. A wave of in-migration is entirely absorbed by firms, which post more vacancies and create more jobs. In fact, the vacancy rate is unaffected by migration, so the number of vacancies grows in proportion to the migration factor. As a result, local workers are unaffected by migration.

If wages respond to in-migration, the DMP model even predicts that in-migration reduces the unemployment rate (figure~\ref{f:dmpwage}). In equation~\eqref{e:ldstd}, the wage $w$ drops with in-migration, so tightness goes up with in-migration. In the textbook DMP model, wages are determined by bargaining. A reason for the wage drop is that migrants have worse outside options than local workers and cannot command the same wages \citep{CP15,A21}. In that case, average wages fall, boosting labor demand. 

\section{Other evidence of migration-induced unemployment}\label{a:other}

Besides the experimental evidence described in section~\ref{s:evidence}, a few papers provide nonexperimental evidence that the arrival of immigrants on a labor market reduces local employment. This appendix briefly discusses these papers.

\citet{C01} uses the 1990 US Census to study the effects of immigration on occupation-specific labor market outcomes. He finds that occupation-specific wages and employment rates are systematically lower in cities with higher relative supplies of workers in each occupation. Hence, immigration over the 1980s reduced wages and employment rates of low-skill natives in gateway cities like Miami and Los Angeles.

\citet{BGH07} use the 1960--2000 US Censuses to examine the correlation between immigration, black wages, and black employment rates. They find that when 100 immigrants arrived in the labor market for a particular skill, 35 black workers were pushed out of employment. They also find that the wage of employed black workers fell, and the incarceration rate of black men increased.

\citet{E15} uses microdata for France, 1990--2002, and finds that in the short run, immigrants increased the unemployment rate of native workers but did not reduce their wages.

\citet{L15} uses provincial panel data to examine the impact of permanent immigration on the unemployment rate in Canada. He finds that in the short run, immigration causes a rise in the unemployment rate. He also finds that this effect disappears in the long run. 

Finally, \citet{D23,D24} uses US state-level data to study the response of labor market tightness to immigration. She finds that the slowdown in immigration between 2017 and 2021 raised the tightness of local labor markets. In 2022 and 2023, the rebound in immigration helped lower labor market tightness (which was inefficiently high).

\section{Elasticities}\label{a:elasticities}

This appendix computes the elasticities of various functions in the model. These elasticities are then used to prove certain propositions in appendix~\ref{a:proofs}. The elasticities are computed using standard results---compiled for instance in \citet[appendix~B]{M26}.

\subsection{Elasticities of matching rates}

The elasticity of the job-finding rate \eqref{e:f} with respect to tightness is $\e^f_{\t} = (1-\h)$. The elasticity of the job-filling rate \eqref{e:q} with respect to tightness is $\e^q_{\t} = -\h$.

\subsection{Elasticity of unemployment rate}

The elasticity of the unemployment rate $u(\t)$, given by \eqref{e:u}, with respect to tightness is:
\begin{equation*}
\e^{u}_{\t} = -\frac{f(\t)}{\l+f(\t)}\cdot \e^f_{\t}.
\end{equation*}
The elasticity simplifies to:
\begin{equation}
\e^{u}_{\t} = - (1-\h) \bs{1-u(\t)}.
\label{e:eut}\end{equation}

The elasticity of $1-u(\t)$ with respect to tightness is:
\begin{equation*}
\e^{1-u}_{\t} = -\frac{u}{1-u}\cdot \e^u_{\t},
\end{equation*}
which simplifies to:
\begin{equation}
\e^{1-u}_{\t} = (1-\h) u(\t).
\label{e:e1-ut}\end{equation}

\subsection{Elasticities of labor supply}

Because labor supply satisfies $l^s(\t) = [1-u(\t)] H$, its elasticity with respect to tightness is the same as the elasticity of $1-u(\t)$:
\begin{equation}
\e^{s}_{\t} = \bp{1-\h} u(\t).
\label{e:est}\end{equation}

The elasticity of labor supply \eqref{e:ls} with respect to migration is simply
\begin{equation*}
\e^{s}_{I} = 1.
\end{equation*}

\subsection{Elasticity of recruiting wedge}

Then the elasticity of $1+\tau$, which is given by \eqref{e:1+tau}, with respect to $\t$ satisfies:
\begin{equation*}
\e^{1+\tau}_{\t} = \e^{q}_{\t} - \e^{q-\l \k}_{\t} =-\h - \frac{q(\t)}{q(\t)-\l \k} (-\h),
\end{equation*}
so it simplifies to
\begin{equation}
\e^{1+\tau}_{\t} = \h\tau(\t).
\label{e:etaut}\end{equation}

\subsection{Elasticities of labor demand}

The elasticity of labor demand \eqref{e:ld} with respect to tightness admits the following expression:
\begin{equation*}
\e^{d}_{\t}= -\frac{1-\a}{\a}\e^{1+\tau}_{\t}.
\end{equation*}
From \eqref{e:etaut}, the elasticity simplifies to
\begin{equation}
\e^{d}_{\t} = -\frac{1-\a}{\a}\cdot \h\tau(\t).
\label{e:edt}\end{equation}

\subsection{Elasticity of welfare}

Finally, I compute the elasticity of local welfare \eqref{e:welfare} with respect to in-migration, allowing not only employment but also wages to respond to migration. I find:
\begin{equation*}
\e^{\Nc}_{I} = \e^{w}_{I} + \e^{N}_{I} + \frac{\a I}{\a I + 1- \a}.
\end{equation*}
Using the expressions for $\e^{N}_{I}$ and $\e^w_{I}=-\b$, and inserting the migration rate $i = I - 1$, I get
\begin{equation}
\e^{\Nc}_{I} = \frac{\a (1+i)}{1 + \a i} - \b - \frac{1-\b/\a}{1-\e^{d}_{\t}/\e^s_{\t}}.
\label{e:ewm}\end{equation}

\subsection{Elasticities with heterogeneous labor}

Equations \eqref{e:phd1} and \eqref{e:pld1} imply that the firm cost shares,
\begin{align*}
s_h &= \hat{w}_h \cdot \frac{P_h^c(\hat{w}_h, \hat{w}_l, X)}{\hat{c}(\hat{w}_h, \hat{w}_l,X)}\\
s_l &= \hat{w}_l \cdot \frac{P_l^c(\hat{w}_h, \hat{w}_l, X)}{\hat{c}(\hat{w}_h, \hat{w}_l,X)}, 
\end{align*}
are the partial elasticities of the unit cost function:
\begin{align}
s_h &= \frac{\hat{w}_h}{c(\hat{w}_h,\hat{w}_l)}\cdot\pd{c(\hat{w}_h,\hat{w}_l)}{\hat{w}_h}\label{e:sh}\\
s_l&= \frac{\hat{w}_l}{c(\hat{w}_h,\hat{w}_l)}\cdot \pd{c(\hat{w}_h,\hat{w}_l)}{\hat{w}_l}.\label{e:sl}
\end{align}

Using the elasticity \eqref{e:etaut}, I compute the elasticity of the effective wage $\hat{w}_{j}(\t_{j})$ with respect to tightness $\t_{j}$:
\begin{equation*}
\e^{\hat{w}_{j}}_{\t_{j}} = \h_{j}\tau_{j}(\t_{j}).
\end{equation*}

From~\eqref{e:phd2} and \eqref{e:pld2} and the cost-share identity in \eqref{e:sl} and \eqref{e:sh}, I then obtain the elasticities of the conditional skill demands with respect to the effective wages:
\begin{align*}
\e^{c,j}_{\hat{w}_{j}} &= -\s s_{-j}\\
\e^{c,j}_{\hat{w}_{-j}} &= \s s_{-j}.
\end{align*}
Substituting into \eqref{e:ldj} yields the demand elasticities with respect to tightness:
\begin{align}
\e^{d,j}_{\t_{j}} &= \h_{j}\tau_{j}(\t_{j}) \bp{1 - \s s_{-j} - \frac{s_{j}}{\a}}\label{e:own}\\
\e^{d,j}_{\t_{-j}} &= \h_{-j}\tau_{-j}(\t_{-j}) s_{-j} \bp{\s - \frac{1}{\a}}.\label{e:cross}
\end{align}

\section{Proofs of the propositions}\label{a:proofs}

This appendix proves the propositions presented in the text. It uses in many places the elasticities computed in appendix~\ref{a:elasticities}.

\subsection{Proof of proposition \ref{p:migration}}

The main step is to compute the elasticity of labor market tightness with respect to the migration factor, denoted $\e^{\t}_{I}$. To do that, I implicitly differentiate the supply-equals-demand condition \eqref{e:solutioni} in elasticities:
\begin{equation*}
\e^{s}_{\t} \cdot \e^{\t}_{I} + 1 = \e^{d}_{\t} \cdot \e^{\t}_{I},
\end{equation*}
where $\e^{s}_{\t}$ and $\e^{d}_{\t}$ are the elasticities of labor supply and demand with respect to tightness, and I used the fact that the elasticity of labor supply with respect to migration simply is $\e^{s}_{I}=1$. I therefore find:
\begin{equation}
\e^{\t}_{I} = \frac{-1}{\e^{s}_{\t}-\e^{d}_{\t}}.
\label{e:eth}\end{equation}

Using the expressions for the elasticities of labor supply and demand $\e^s_{\t}$ and $\e^d_{\t}$ given by equations \eqref{e:est} and \eqref{e:edt}, the elasticity becomes:
\begin{equation*}
\e^{\t}_{I} = - \frac{1}{(1-\h) u(\t) + \frac{1-\a}{\a}\cdot \h\tau(\t)}. 
\end{equation*}
This is just the expression given by \eqref{e:thetai}.

The elasticity of tightness with respect to migration is negative. So with in-migration, labor market tightness drops, the unemployment rate $u(\t)$ increases, and as the local labor force is unchanged, the number of local workers in unemployment rises and the number $N$ of local workers in employment drops. Aggregate employment can be read off the labor demand, $L = L^d(\t)$, so it rises when tightness drops.

\subsection{Proof of proposition \ref{p:wage}}

By implicitly differentiating $L^s(\t,I) = L^d(\t,I)$ in elasticities, I obtain:
\begin{equation*}
\e^{s}_{\t} \cdot \e^{\t}_{I} + 1 = \e^{d}_{\t} \cdot \e^{\t}_{I} + \frac{\b}{\a},
\end{equation*}
where I used the facts that the elasticity of the wage \eqref{e:wage} with respect to migration is $\e^w_{I} = -\b$, and the elasticity of labor demand \eqref{e:ld} with respect to the wage is $\e^{d}_w = -1/\a$. 

Thus, the elasticity of tightness with respect to migration is
\begin{equation*}
\e^{\t}_{I} = -\bp{1-\frac{\b}{\a}} \cdot \frac{1}{\e^{s}_{\t}-\e^{d}_{\t}},
\end{equation*}
which, just as in the proof of proposition \ref{p:migration}, can be written:
\begin{equation*}
\e^{\t}_{I} = -\bp{1-\frac{\b}{\a}} \cdot \frac{1}{(1-\h) u(\t) + \frac{1-\a}{\a}\cdot \h\tau(\t)}. 
\end{equation*}

The elasticity of tightness with respect to migration is negative. So with in-migration, labor market tightness drops, the unemployment rate $u(\t)$ increases, and as the local labor force is unchanged, the number of local workers in unemployment rises and the number $N$ of local workers in employment drops.

Aggregate employment can be read off the labor demand, $L = L^d(\t,I)$. The labor demand is decreasing in tightness $\t$ but increasing in migration $I$ (as migration reduces wages). So labor demand rises when tightness drops and in-migration occurs, which implies that aggregate employment grows with in-migration.

\subsection{Proof of proposition \ref{p:businesscycle}}

The elasticity of tightness with respect to productivity, denoted $\e^{\t}_a$, is obtained by implicitly differentiating the supply-equals-demand equation in elasticity:
\begin{equation*}
\e^{s}_{\t} \cdot \e^{\t}_{a} = \e^{d}_{\t} \cdot \e^{\t}_{a} + \e^{d}_a.
\end{equation*}
Given the expression \eqref{e:lda}, the elasticity of the labor demand with respect to productivity is $\e^d_a = \g/\a$. Thus, the elasticity of tightness with respect to productivity is
\begin{equation*}
\e^{\t}_a = \frac{\g}{\a} \cdot \frac{1}{\e^{s}_{\t}-\e^{d}_{\t}}.
\end{equation*}
Using the expressions for the elasticities of labor supply and demand $\e^s_{\t}$ and $\e^d_{\t}$ given by equations \eqref{e:est} and \eqref{e:edt}, the elasticity becomes:
\begin{equation*}
\e^{\t}_a = \frac{\g}{\a} \cdot \frac{1}{(1-\h) u(\t) + \frac{1-\a}{\a}\cdot \h\tau(\t)}.
\end{equation*}
This is just the expression given by \eqref{e:thetaa}.

With wage rigidity, $\g>0$, so that $\e^{\t}_a>0$: when productivity drops, tightness falls. Accordingly, the job-finding rate $f(\t)$ drops, unemployment rate $u(\t)$ rises, the job-filling rate $q(\t)$ rises, and the recruiter-producer ratio $\tau(\t)$ declines.

\subsection{Proof of proposition \ref{p:stealing}}

From the elasticity $\e^{\t}_{I}$ of labor market tightness with respect to migration, given by \eqref{e:thetaiwage}, I infer the response of local employment \eqref{e:ni} to migration. The response is measured by the elasticity of local employment with respect to the migration factor $I$. That elasticity is given by $\e^{N}_{I} = \e^{1-u}_{\t} \cdot \e^{\t}_{I}$, where the elasticity of $1-u(\t)$ is given by equation \eqref{e:e1-ut}. Hence, the elasticity becomes:
\begin{equation*}
\e^{N}_{I}(\t) = - \bp{1-\frac{\b}{\a}} \cdot \frac{1}{1+\frac{1-\a}{\a} \cdot \frac{\h}{1-\h}\cdot \frac{\tau(\t)}{u(\t)}}. 
\end{equation*}
This is just the expression given by \eqref{e:elh2}.

Since $\tau(\t)$ is increasing with $\t$ while $u(\t)$ is decreasing with $\t$, the amplitude of $\e^{N}_{I}(\t)$ is decreasing with tightness. When $\t\to 0$, $\tau(\t)\to 0$, so $\e^{N}_{I} \to -(1-\b/\a)$. When tightness is efficient, $\h\tau(\t) = (1-\h) u(\t)$, so $\e^{N}_{I} \to -(\a - \b)$. When $\t\to \bar{\t}$, $\tau(\t)\to \infty$, so $\e^{N}_{I} \to 0$.

\subsection{Proof of proposition \ref{p:welfare}}

As a first step, I assess whether any in-migration might ever improve welfare. To do that, I determine whether the elasticity $\e^{\Nc}_{I}$ might ever be positive at $i = 0$. That is, I compute the effect of an infinitesimal wave of in-migration on welfare.

Setting $i = 0$ in \eqref{e:ewm} and using \eqref{e:edt} and \eqref{e:est}, I find:
\begin{equation}
\e^{\Nc}_{I} =\bp{\a - \b } \bs{1 - \frac{1}{\a+(1-\a)\frac{\h}{1-\h}\frac{\tau(\t)}{u(\t)}}}.
\label{e:ewm1}\end{equation}

When $\t \to \bar{\t}$, $\tau(\t) \to \infty$, so $\e^{\Nc}_{I} \to\a-\b>0$. Thus, when the labor market is at its tightest (at which point all workers are recruiters), then in-migration is desirable.

When $\t \to 0$, $\tau(\t) \to 0$, so $\e^{\Nc}_{I} \to (\a-\b)(1-1/\a)<0$. Hence, when the labor market is at its slackest (at which point all workers are unemployed), then in-migration is undesirable.

Given that $\tau/u$ is strictly increasing in $\t\in(0,\bar{\t})$, $\e^{\Nc}_{I}$ is continuous and strictly increasing in $\t$, and there exists a unique $\t_{i}\in(0,\bar{\t})$ at which $\e^{\Nc}_{I} = 0$, and in-migration improves welfare at any $\t>\t_{i}$ while in-migration reduces welfare at any $\t<\t_{i}$.

Solving for $\e^{\Nc}_{I}(\t) = 0$ with \eqref{e:ewm1}, I obtain
\begin{equation}
\frac{\h}{1-\h}\cdot\frac{\tau(\t)}{u(\t)} = 1.
\label{e:efficiency1}\end{equation}
But this condition is just the efficiency condition in the model \citep[lemma 1]{MS19}. The tightness $\t_{i}$ is just the tightness that maximizes the number of producers and consumption for a given labor force---the efficient tightness $\t^*$. 

It is easy to see why. Consumption is determined by the number of producers, 
\begin{equation*}
P = \frac{L}{1+\tau(\t)} = \frac{1-u(\t)}{1+\tau(\t)} H. 
\end{equation*}
Maximizing the number of producers is the same as maximizing the share of labor that is productive,
\begin{equation*}
\frac{1-u(\t)}{1+\tau(\t)}.
\end{equation*}
The elasticity of $1-u(\t)$ with respect to tightness is given by equation~\eqref{e:eut}. The elasticity of $1+\tau(\t)$ with respect to tightness is given by equation~\eqref{e:etaut}. The number of producers is maximized for a given labor force when its elasticity with respect to tightness is 0, or 
\begin{equation*}
\h \tau(\t)= (1-\h) u(\t). 
\end{equation*}

\subsection{Proof of proposition \ref{p:optimal}}

The question boils down to finding the migration rate $i^*$ such that the elasticity $\e^{\Nc}_{I} = 0$. At that migration rate, local welfare is maximized so migration is optimal.

Using \eqref{e:ewm}, the optimality condition $\e^{\Nc}_{I} = 0$ becomes
\begin{align*}
\frac{\a (1+i)}{1 + \a i} - \b & = \frac{\a-\b}{\a-\a \e^{d}_{\t}/\e^s_{\t}}.
\end{align*}
Using the expressions \eqref{e:est} and \eqref{e:edt} for the elasticities of demand and supply with respect to tightness, I rewrite the condition as
\begin{equation}
\frac{\a (1+i)}{1 + \a i} - \b = \frac{\a-\b}{\a+(1-\a) \frac{\h}{1-\h} \cdot\frac{\tau(\t)}{u(\t)}}.
\label{e:istar}\end{equation}

Since tightness is itself a function of the migration rate $i$, the optimal in-migration is the solution to a fixed-point problem. The optimal in-migration rate $i^*>0$ solves the fixed-point equation \eqref{e:istar}. The left-hand side of the equation is an increasing function of $i$. At $i=0$, the left-hand side is $\a-\b$, so at optimum in-migration $i^*>0$, the left-hand side is strictly greater than $\a-\b$. The implication is that at optimum in-migration, the right-hand side is also strictly greater than $\a-\b$. This requires the denominator of the right-hand side to be less than 1, and accordingly
\begin{equation*}
\frac{\h}{1-\h} \cdot\frac{\tau(\t)}{u(\t)}<1.
\end{equation*}
Hence, labor market tightness must be strictly less than the efficient tightness $\t^*$, which is defined by \eqref{e:efficiency1}. Thus, at optimum in-migration, the labor market is inefficiently slack.

\subsection{Proof of proposition \ref{p:heterogeneity}}

Just like in the simple case with homogeneous labor, I implicitly differentiate the supply-equals-demand equations \eqref{e:equilibrium1} and~\eqref{e:equilibrium2} to obtain the elasticities of tightnesses with respect to low-skill in-migration. 

I consider a change in the low-skill labor force, $H_l$, which represents a low-skill in-migration shock. The linear system of elasticity equations can be written:
\begin{equation}
\begin{bmatrix}
\e^{d,l}_{\t_l} - \e^{s,l}_{\t_l} & \e^{d,l}_{\t_h} \\
\e^{d,h}_{\t_l} & \e^{d,h}_{\t_h} -\e^{s,h}_{\t_h} 
\end{bmatrix}
\begin{bmatrix} 
\e^{\t_l}_{H_l} \\ 
\e^{\t_h}_{H_l} 
\end{bmatrix}
=
\begin{bmatrix} 
1 \\ 
0 
\end{bmatrix},
\label{e:system}\end{equation}
where $\e^{d,j}_{\t_j}$ is the elasticity of the demand for skill $j$ with respect to the tightness in skill market $j$ and $\e^{s,j}_{\t_{j}}$ is the elasticity of the supply of skill $j$ with respect to the tightness in skill market $j$. The vector on the right-hand side contains the elasticities of the labor supplies with respect to $H_l$: 1 for the supply of skill $l$ and 0 for the supply of skill $h$.

By Cramer's rule, the responses to a low-skill in-migration shock in which $H_l$ rises are
\begin{align}
\e^{\t_l}_{H_l} = \frac{\e^{d,h}_{\t_h} - \e^{s,h}_{\t_h}}{\D}\label{e:etlhl}\\
\e^{\t_h}_{H_l} = -\frac{\e^{d,h}_{\t_l}}{\D}\label{e:ethhl}.
\end{align}
where $\D$ is the determinant of the matrix in \eqref{e:system}:
\begin{equation}
\D = \bs{\e^{d,h}_{\t_h} -\e^{s,h}_{\t_h}}\bs{\e^{d,l}_{\t_l} - \e^{s,l}_{\t_l}} - \e^{d,l}_{\t_h} \e^{d,h}_{\t_l}.
\label{e:delta}\end{equation}

Substituting the elasticities \eqref{e:est}, \eqref{e:own}, and \eqref{e:cross} yields
\begin{align*}
\D &= (1-\h_l)u_l\cdot(1-\h_h)u_h - (1-\h_l)u_l\cdot \h_h\tau_h\bp{1-\s s_l -\frac{s_h}{\a}} \\
& - (1-\h_h)u_h\cdot \h_l\tau_l\bp{1-\s s_h -\frac{s_l}{\a}} - \h_l\tau_l\cdot \h_h\tau_h\cdot \frac{(1-\s)(1-\a)}{\a}.
\end{align*}
Since $\s> 1$, the last term is positive. In addition,
\begin{equation*}
1-\s s_{-j}-\frac{s_{j}}{\a} \le 1-s_{-j}-\frac{s_{j}}{\a} = -s_{j}\frac{1-\a}{\a}<0,
\end{equation*}
so the two middle terms are positive. The first term is positive. Therefore $\D>0$.

The own-elasticity \eqref{e:own} is unambiguously negative. Since $\s> 1$, then $\s s_{-j}> s_{-j}$. Since $\a<1$, $s_{j}/\a > s_{j}$. And as $s_{j}+s_{-j}=1$, then in this case $\s s_{-j} + s_{j}/\a > 1$, so the own elasticity is strictly negative. The cross-elasticity \eqref{e:cross} inherits its sign from $\s - 1/\a$.

Since $\D>0$, $\e^{d,h}_{\t_h}<0$, and $\e^{s,h}_{\t_h}>0$, the own-tightness response \eqref{e:etlhl} is negative. Low-skill tightness therefore falls with low-skill in-migration, and the low-skill unemployment rate rises. The cross-tightness response is given by \eqref{e:ethhl}. The cross-elasticity $\e^{d,h}_{\t_l}$ has the sign of $\s-1/\a$. Because $\D>0$, the elasticity of high-skill tightness $\e^{\t_h}_{H_l}$ has the opposite sign. Hence, if $\s>1/\a$, high-skill tightness falls and high-skill unemployment rises with low-skill in-migration. If $\s<1/\a$, high-skill tightness rises and high-skill unemployment falls with low-skill in-migration.

\section{Existence and uniqueness of the model solution}\label{a:solution}

This appendix proves that the solution of the model exists and is unique whether labor is homogeneous or heterogeneous.

\subsection{Homogeneous labor}

I first consider the model with homogeneous labor described in section~\ref{s:model}. I establish the existence and uniqueness of the solution by introducing the excess labor demand function:
\begin{equation}
z(\t,H) = L^d(\t) - L^s(\t,H).
\label{e:z}\end{equation}
With this function, the solution of the model \eqref{e:solution} simply is
\begin{equation*}
z(\t,H) = 0.
\end{equation*}

Labor demand $L^d(\t)$ is strictly decreasing in tightness, from $L^d(0)>0$ when $\t=0$ to $L^d(\bar{\t}) = 0$ when $\t = \bar{\t}$. At the same time, labor supply $L^s(\t)$ is strictly increasing in tightness, starting from $L^s(0,H) = 0$ when $\t =0$. Moreover, both $L^d$ and $L^s$ are continuous in tightness. 

Thus, the excess demand function $z$ is continuous in $\t$, and it declines from $z(0,H) = L^d(0)>0$ to $z(\bar{\t},H) = - L^s(\bar{\t},H)<0$ when $\t$ increases from $0$ to $\bar{\t}$. Bolzano's theorem then says that equation $z(\t,H) = 0$ has a solution on $(0,\bar{\t})$. Moreover, since the excess demand function is strictly decreasing, the solution is unique.

\subsection{Heterogeneous labor}

Next I turn to the model with heterogeneous labor introduced in section~\ref{s:heterogeneity}. The excess labor demands in that case are
\begin{equation*}
Z_{j}(\t_l, \t_h) = L_{j}^d(\t_l, \t_h) - L_{j}^s(\t_{j}, H_{j}),
\end{equation*}
where $j =l$ for low-skill labor and $j =h$ for high-skill labor.

The excess demand mapping $(Z_l, Z_h)$ is continuous on $(0, \bar{\t}_l) \times (0, \bar{\t}_h)$. As $\t_{j} \to 0$, $L_{j}^s \to 0$ while $L_{j}^d>0$, so $Z_{j} > 0$. As $\t_{j} \to \bar{\t}_{j}$, $\tau_{j} \to \infty$, so $\hat{w}_{j} \to \infty$, and $L_{j}^d \to 0$ while $L_{j}^s>0$, so $Z_{j} < 0$. By the continuity of $(Z_l, Z_h)$ and the Poincare-Miranda theorem, there exists a point $(\t_l, \t_h)$ at which $Z_h(\t_l, \t_h) = Z_l(\t_l, \t_h) = 0$.

Turning to uniqueness, it is convenient to work with the excess demands in logarithms:
\begin{equation*}
G_{j}(\t_l,\t_h) = \ln L_{j}^d(\t_l, \t_h) - \ln L_{j}^s(\t_{j}, H_{j}).
\end{equation*}
The solutions of $G_l = G_h = 0$ are exactly the solutions of $Z_l = Z_h = 0$. The advantage of the logarithmic formulation is that its Jacobian is the elasticity matrix that already appears in \eqref{e:system}. Indeed, $\pd{G_{j}}{\t_k} = \bs{\e^{d,j}_{\t_k} - \e^{s,j}_{\t_k}}/\t_k$, with $\e^{s,j}_{\t_k} = 0$ for $j \neq k$, so the Jacobian of $G$ is $M D^{-1}$, where $M$ is the matrix in \eqref{e:system} and $D$ is the diagonal matrix with entries $\t_l$ and $\t_h$. This holds at every point of the rectangle, not only at a solution.

I now apply the univalence theorem of \citet{GN65}: a differentiable mapping whose Jacobian is a P-matrix at every point of a rectangular region is injective on that region. The relevant mapping is $-G$ rather than $G$, because a P-matrix has positive principal minors while the own effects here are negative. The Jacobian of $-G$ is $-M D^{-1}$, and since $D^{-1}$ is diagonal with positive entries, each principal minor of $-M D^{-1}$ is a positive multiple of the corresponding principal minor of $-M$. It therefore suffices that $-M$ be a P-matrix.

For a $2\times 2$ matrix, that requires positive diagonal entries and a positive determinant. The diagonal entries of $-M$ are $\e^{s,j}_{\t_{j}} - \e^{d,j}_{\t_{j}}$, which are positive because labor supply is increasing in tightness and, as established in the proof of proposition~\ref{p:heterogeneity}, the own elasticity of labor demand \eqref{e:own} is negative. The determinant of $-M$ equals $\D$, which the same proof shows to be positive. Both statements hold pointwise in $(\t_l,\t_h)$, so $-M$ is a P-matrix on the entire rectangle.

Hence, the mapping $-G$ is injective on $(0, \bar{\t}_l) \times (0, \bar{\t}_h)$---any two points of the open rectangle lie in a common closed subrectangle---so $G$ vanishes at most once. Together with the existence result above, the model therefore admits exactly one solution.

\section{Calibration of the model}\label{a:calibration}

This appendix calibrates the model simulated in section~\ref{s:illustration} to the US labor market. The calibrated values of the model parameters are summarized in table~\ref{t:calibration}.

\subsection{Concavity of the production function} 

Before it started falling in the 1990s, the US labor share was hovering around 65\% \citep[figure~1]{ADK20}. The global labor share was also at 65\% in 1980 \citep[figure~1]{KN14}. Accordingly, I set the concavity parameter in the production function \eqref{e:production} to $\a=0.35$, which produces a labor share of $1-\a = 65\%$. Targeting a lower labor share would require a higher $\a$, which would generate more migration-induced unemployment (as shown in proposition~\ref{p:stealing}).

\subsection{Rigidity of real wages} 

I set the rigidity parameter in the wage function \eqref{e:wageGeneral} to $\g = 0.3$, as in \citet[table~1]{M12}. Such a calibration produces an elasticity of real wages with respect to productivity of $1-\g = 0.7$, which is in line with the findings by \citet[table~6]{HSV13} and is low enough to generate realistic business cycles \citep[section~4B]{M12}.

\subsection{Migration elasticity of real wages} 

The empirical response of wages to migration is captured through the elasticity $\b$ in the wage norm \eqref{e:wageGeneral}. Available evidence suggests that migration does not generally have significant effects on local wages \citep{FH95,RR07,BK15}. So as a baseline I simply set $\b=0$.

\subsection{Recruiting cost} 

I set the recruiting cost to $\k = 1$, following \citet[section~2C]{MS24}. This calibration is based on the evidence provided by \citet{GMV18} and \citet{MS21b}. It implies that it takes 1 worker to service a job vacancy.

\subsection{Job-separation rate} 

I set the quarterly job-separation rate to $\l = 0.097$, which is its average value in the United States between 1951 and 2019 \citep[appendix~B2]{MS21b}.

\subsection{Matching elasticity} 

I set the elasticity parameter in the matching function \eqref{e:matching} so that the elasticity of the Beveridge curve is $1$, as it is in the United States \citep[section~2E]{MS24}. The matching elasticity $\h$ is related to the elasticity of the Beveridge curve $\e$ and unemployment rate $u$ by 
\begin{equation*}
\e = \frac{1}{1-\h}\bp{\h + \frac{u}{1-u}},
\end{equation*}
as shown by \citet[equation (12)]{MS21b}. Since $\h$ is fixed, the Beveridge curve is not exactly isoelastic, as $\e$ varies a little bit when $u$ varies. Hence, I target a Beveridge elasticity of $1$ when the unemployment rate is efficient. The above equation can be rewritten as
\begin{equation*}
\h = \frac{1}{1+\e} \bp{\e - \frac{u}{1 - u}}.
\end{equation*}
The average value of the efficient unemployment rate in the United States over 1951--2019 is $u^* = 4.3\%$ \citep[section 3A]{MS24}. I plug $u = 4.3\%$ and $\e=1$ in the above equation. I obtain $\h = 0.48$. This calibration is very close to the value $\h = 0.5$ which is standard in the literature \citep{MP94,PP01,S04,HM08}.

\subsection{Matching efficacy} 

I calibrate the efficacy parameter in the matching function \eqref{e:matching} so that the location of the Beveridge curve is the same as in US data. The location of the Beveridge curve determines the efficient unemployment rate $u^*$, so the calibration ensures that the efficient unemployment rate is the same in the model as in the data. The average value of the efficient unemployment rate in the United States over 1951--2019 is $u^* = 4.3\%$, while the efficient tightness is $\t^* = 1$ \citep[section 3A]{MS24}. The efficient unemployment rate, efficient tightness, and matching efficacy are themselves related by
\begin{equation*}
\m = \frac{\l}{(\t^*)^{1-\h}} \cdot \frac{1 - u^*}{u^*}.
\end{equation*}
This equation is obtained by reshuffling \eqref{e:u}. Setting tightness and unemployment to their efficient levels, $\t^* = 1$ and $u^* = 4.3\%$, and setting $\l = 0.097$, this equation yields $\m = 2.16$.

\subsection{Real wage} 

I calibrate the real-wage level so that the labor market is efficient for a productivity level of 1---which is just a normalization. Through the labor demand \eqref{e:ld}, the wage level is related to various parameters by
\begin{equation*}
\o = \frac{(1 - \a) a^{\g}}{(1 - u^*)^\a} \cdot \bs{1 + \frac{\k\l}{\m \cdot (\t^*)^{-\h} - \k\l}}^{\a-1}.
\end{equation*}
With $\a = 0.35$, $a=1$, $u^* = 0.043$, $\k=1$, $\l = 0.097$, $\m = 2.16$, $\t^* =1$, the equation gives $\o = 0.64$.

\begin{figure}[p]
\subcaptionbox{Migration elasticity of local employment\label{f:elhwage}}{\normalgraphic[17]{\pdf}}\hfill
\subcaptionbox{Migration elasticity of local welfare\label{f:ewhwage}}{\normalgraphic[18]{\pdf}}\fspace
\subcaptionbox{Optimal in-migration policy\label{f:istarwage}}{\normalgraphic[19]{\pdf}}\hfill
\subcaptionbox{Optimal labor market tightness\label{f:thetastarwage}}{\normalgraphic[20]{\pdf}}
\caption{Model behavior and optimal policy for various wage responses to migration}
\note{The figure is obtained by simulating the labor market model under the calibration presented in table~\ref{t:calibration}. The pink lines represent the efficient location of the labor market. The purple lines describe how the results change for different migration elasticities of real wages, from $\b=0$ (unresponsive wages) to $\b = 0.15$ (somewhat responsive wages). Panel A plots $\e^{N}_{I}(\t)$, computed using \eqref{e:elh2}. Panel B plots $\e^{\Nc}_{I}(\t)$, computed using \eqref{e:ewm}. Tightness $\t(a,i)$ is a function of productivity $a$ and migration rate $i$ given by \eqref{e:solution}. Panel C and D plot $i^*(a)$ and $\t(a,i^*(a))$ against initial tightness $\t(a,0)$, computed by solving \eqref{e:istar}.}
\end{figure}

\section{Simulations for different wage responses to migration}\label{a:wage}

This appendix explores the behavior of the model for different wage responses to migration. It simulates the model for all values of the wage-flexibility parameter $\b$ between $\b = 0$ and $\b = 0.15 < \a$, which corresponds to a somewhat flexible wage.

A first observation is that the elasticity of local employment with respect to migration shrinks to zero as the response of wages to migration grows (figure~\ref{f:elhwage}). When real wages fall in response to in-migration, then in-migration stimulates not only labor supply but also labor demand, so tightness falls less when in-migration occurs, meaning that local employment falls less and the elasticity is closer to 0.

A second observation is that the optimal amount of in-migration is larger when wages fall in response to in-migration (figure~\ref{f:istarwage}). This result might seem surprising, given that the migration elasticity of welfare is less positive when wages fall (figure~\ref{f:ewhwage}). But this is just a marginal result, which determines the sign but not the optimal amount of migration. Even when wages respond to migration, optimal migration should continue to bring tightness close to 1 (figure~\ref{f:thetastarwage}). At the same time, the effect of migration on tightness falls when wages respond more (figure~\ref{f:elhwage}). Accordingly, since migration is less potent when wages respond more, a larger in-migration rate is required to bring tightness to the vicinity of its efficient level of 1.

\end{document}